\documentclass[fleqn,12pt,a4paper]{article}
\usepackage[utf8]{inputenc}
\usepackage{makeidx}
\usepackage{graphicx}
\usepackage{amsmath,amssymb} 
\usepackage{amsfonts}
\usepackage{float}
\usepackage{tikz}
\usepackage{color}
\usepackage{xcolor}
\usepackage[most]{tcolorbox}
\usepackage{framed}
\usepackage{adjustbox}
\usepackage{pgfplots}
\usepackage{subfig}
\usepackage[makeroom]{cancel}
\usepackage{xspace}
\usepackage[hyphens]{url}
\usepackage[square, numbers]{natbib}
\usepackage{hyperref}

\pgfplotsset{compat=1.14}

\newcommand{\ie}{i.e.\@\xspace}
\newcommand{\Ie}{I.e.\@\xspace}

\newcommand{\etal}{et al.\@\xspace}

\usetikzlibrary{arrows}
\usetikzlibrary{shapes.geometric}
\usetikzlibrary{shapes.misc, positioning}
\usetikzlibrary{shapes.multipart, shapes.arrows}

\tikzset{
	treenode/.style = {align=center, inner sep=0pt, text centered,
		font=\sffamily},
	tochild/.style={draw,-latex},
	toparent/.style={draw,latex-},
	whitenode/.style = {treenode, circle, black, draw=black, text width=1.5em, very thick}, 
	block/.style= {draw, rectangle, minimum width=3cm,minimum height=1cm},
}

\tcbset{textmarker/.style={%
		enhanced,
		parbox=false,boxrule=0mm,boxsep=0mm,arc=0mm,
		outer arc=0mm,left=6mm,right=3mm,top=7pt,bottom=7pt,
		toptitle=1mm,bottomtitle=1mm,oversize}}

\newtcolorbox{exeBox}{textmarker, borderline west={6pt}{0pt}{gray}, colback=gray!10!white}
\newtcolorbox{blueBox}{textmarker, borderline west={6pt}{0pt}{blue}, colback=blue!10!white}
\newcommand{\exercise}[2]{\begin{exeBox} \textbf{Exercise #1:} #2 \end{exeBox}}
\newcommand{\blueteam}[1]{\begin{blueBox} \textbf{Blue-Team:} #1 \end{blueBox}}
\newcommand{\gitlink}[1]{\noindent$\to$ \url{https://github.com/Kayzaks/HackingNeuralNetworks/tree/master/#1}}

\begin{document}

\title{Hacking Neural Networks:\\A Short Introduction}

\author{Michael Kissner}

\date{v1.03 (November 22, 2019)}

\maketitle

\begin{abstract}
A large chunk of research on the security issues of neural networks is focused on adversarial attacks. However, there exists a vast sea of simpler attacks one can perform both against and with neural networks. In this article, we give a quick introduction on how deep learning in security works and explore the basic methods of exploitation, but also look at the offensive capabilities deep learning enabled tools provide. All presented attacks, such as backdooring, GPU-based buffer overflows or automated bug hunting, are accompanied by short open-source exercises for anyone to try out.  
\end{abstract}

\clearpage

\tableofcontents

% \cite{Zhang:2018} - Deep Learning in Mobile and Wireless Networking: A Survey
% \cite{Curtin:2019} - Detecting DGA domains with recurrent neural networks and side information

% Exercise ideas:
%  0-0 - Train the model yourself
%  X-X - Reinforcement Learning Pen Testing

\section{Introduction}

\textcolor{black}{\textbf{Disclaimer: This article and all the associated exercises are for educational purposes only.}}

\medskip

% Intro on how to create a mini-version of a model to test on and that a more elaborate version might work on the actual model?

When one looks for information on exploiting neural networks or using neural networks in an offensive manner, most of the articles and blog posts are focused on adversarial approaches and only give a broad overview of how to actually get them to work. These are certainly interesting and we will investigate what "adversarial" means and how to perform simplified versions of these attacks, but our main focus will be on all the other methods that are easy to understand and exploit.
 
Sadly, we can't cover everything. The topics we do include here were chosen because we feel they provide a good basis to understand more complex methods and allow for easy to follow exercises. We begin with a quick introduction to neural networks and move on to progressively harder subjects. The goal is to point out security issues, demystify some of the daunting aspects of deep learning and show that its actually really easy to get started and mess around with neural networks.

\medskip

\noindent \textbf{Who its for:} This article is aimed at anyone that is interested in deep learning from a security perspective, be it the defender faced with a sudden influx of applications utilizing neural networks, the attacker with access to a machine running such an application or the CTF-participant who wants to be prepared. 

\medskip

\noindent\textbf{How to setup:} To be able to work on the exercises, we need to prepare our environment. For speed, it is advisable to use a computer with a modern graphics card. We will also need:
\begin{enumerate}
	\item \textbf{Python and pip:} Download and install Python3 and its package installer pip using a package manager or directly from the website \url{https://www.python.org/downloads/}. 
	\item \textbf{Editor:} An editor is required to work with the code, preferably one that allows code highlighting for Python. Vim/Emacs will do. For reference, all exercises were prepared using Visual Studio Code \url{https://code.visualstudio.com/docs/python/python-tutorial}.
	\item \textbf{Keras:} Installing Keras can be tricky. We refer to the official installation guide at \url{https://keras.io/#installation} and suggest TensorFlow as a backend (using the GPU-enabled version, if one is available on the machine) as it is the most prevalent in industry \cite{He:2019}.
	\item \textbf{NumPy, SciPy and scikit-image:} NumPy and SciPy are excellent helper packages, which are used throughout all exercises. Following the official SciPy instructions should also install NumPy \url{https://www.scipy.org/install.html}. We will also need to install scikit-image for image loading and saving: \url{https://scikit-image.org/docs/stable/install.html}.
	\item \textbf{NLTK:} NLTK provides functionalities for natural language processing and is very helpful for some of the exercises. \url{https://www.nltk.org/install.html}.
	\item \textbf{PyCuda:} PyCuda is required for the GPU-based attack exercise. If no nVidia GPU is available on the machine, this can be skipped \url{https://wiki.tiker.net/PyCuda/Installation}.
\end{enumerate}

\clearpage
\medskip\noindent\textbf{What else to know: }

\begin{itemize}
	\item The exercises convey half of the content and it is recommended to at least look at them and their solutions. While it is helpful to have a good grasp of python, a lot of the exercises can be solved with basic programming knowledge and some tinkering.
	\item All code is based on open-source deep learning code, which was modified to work as an exercise. This is somewhat due to laziness, but mainly because this is the actual process a lot developers follow: Find a paper that seems to solve the problem, test out the reference implementation and tweak it until it works. Where applicable, a reference is given in the code itself to what it is based on.
	\item This is meant as a living document. Should an important reference be missing or some critical error still be present in the text, please contact the author.
\end{itemize}

\subsection{Quick Guide to Neural Networks}

In this section, we will take a quick dive into how and why neural networks work, the general idea behind learning and everything we need to know to move on to the next sections. We'll take quite a different route compared to most other introductions, focusing on intuition and less on rigor. If you are familiar with the overall idea of deep learning, feel free to skip ahead. As a better alternative to this introduction or as a supplement, we suggest watching 3Blue1Brown's YouTube series \cite{Sanderson:2017} on deep learning.

Let's take a look at a single neuron. We can view it as a simple function which takes a bunch of inputs $x_1, \cdots, x_n$ and generates an output $f(\vec{x})$. Luckily, this function isn't very complex:
\begin{equation}\label{eq:ZZ}
z(\vec{x}) = w_1 x_1 + \cdots + w_n x_n + b \;\;\; ,
\end{equation}
which is then put through something called an activation function $a$ to get the final output
\begin{equation}
f(\vec{x}) = a(z(\vec{x})) \;\;\; .
\end{equation}

\begin{figure}[H]
	\centering
	\begin{tikzpicture}
	%\draw[very thin,color=gray!60] (-6.1,2.0-0.1) grid (-6.0+3.2,2.0+2.2);
	\draw[very thick,->] (-9,2.0) -- (-3,2.0+0) node[right] {$x$};
	\draw[very thick,->] (-6,2.0-0.2) -- (-6.0+0,2.0+2.6) node[above] {ReLU};
	\draw[color=red] plot[domain=-8.6:-3.4,id=fx1] function{x<=-6 ? 2.05 : x+8.05}; 
	
	\end{tikzpicture}
	\caption{The rectified linear unit.} \label{fig:ReLU}
\end{figure}
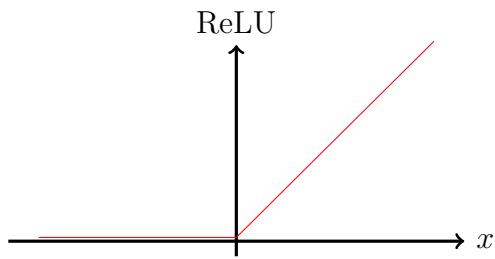

All the inputs $x_i$ are multiplied by the values $w_i$, to which we refer to as weights, and added up together with a bias $b$. The following activation function $a(\cdot)$ basically acts as a gate-keeper. One of the most common such activation functions is the ReLU \cite{Hahnloser:2000}\cite{Glorot:2011}, which has been, together with its variants, in wide use since 2011. ReLU is just the $a(\cdot) = \max(0, \cdot)$ function, which sets all negative outputs to $0$ and leaves positive values unchanged, as can be seen in Figure \ref{fig:ReLU}.

In all we can thus write our neuron simply as
\begin{equation}\label{eq:singleneuron}
f(\vec{x}) = \max(0, w_1 x_1 + \cdots + w_n x_n + b) \;\;\; .
\end{equation}

\begin{figure}[H]
	\centering
		\begin{tikzpicture}
		\draw[very thin,color=gray!60] (-8.6,2.0-0.1) grid (-3.4,2.0+2.2);
		\draw[very thick,->] (-9,2.0) -- (-3,2.0+0) node[right] {$x$};
		\draw[very thick,->] (-6,2.0-0.2) -- (-6.0+0,2.0+2.6) node[above] {$h(x)$};
		\draw[color=red] plot[domain=-8.6:-3.4,id=fx2] function{x < -7 ? 2.05 : x<=-6 ? (x+7) + 2.05 : x<=-5 ?  -(x+5) + 2.05 : 2.05};
		
		\end{tikzpicture}
	\caption{A single hat function.} \label{fig:HatFunction}
\end{figure}
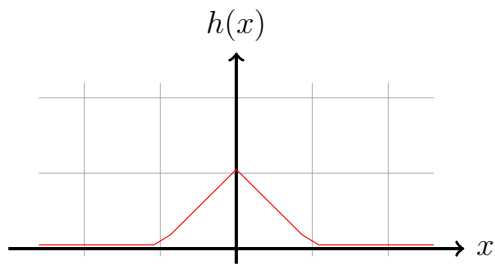

This is admittedly quite boring, so we will try to create something out of it. Neurons can be connected in all sorts of ways. The output of one neuron can be used as the input for another. This allows us to create all sorts of functions by connecting these neurons. As an example, we will use the hat function, which has a small triangular hat at some point and is $0$ almost everywhere else, as shown in Figure \ref{fig:HatFunction}.

We can write down the equation for this function as follows:
\begin{equation}\label{eq:hatfunction}
	h_{i} = \max\left(0,1 -\max\left(0, \frac{c_{i} - x}{c_{i} - c_{i-1}}\right) - \max\left(0, \frac{x - c_{i}}{c_{i+1} - c_{i}}\right)\right) \;\;\; .
\end{equation}

\begin{figure}[H]
	\centering
		\begin{tikzpicture}

		\node (E) at (5, 0)  {Output}  
		child[grow=left]
		{ 
			node [whitenode] at (0, 0.0) {} edge from parent[toparent]
			child[grow=left]
			{ 
				node [whitenode] at (0, 0.5) {} edge from parent[toparent] 
				child[grow=left]
				{
					node (A) at (0, -0.5) {Input} edge from parent[toparent]
				}
			}
			child[grow=left]{ node (B) [whitenode] at (0, -0.5) {} edge from parent[toparent] }
		};
		
		\draw[tochild, black] (A) -- (B);	
				
		\draw [dashed, red] (1.1,0.95) -- (4.2,0.95) -- (4.2,-0.95) -- (1.1,-0.95) -- (1.1,0.95);
		\node [red] at (3.8,-0.6) {$h(x)$};

		\end{tikzpicture}
	\caption{A hat function $h(x)$ represented by 3 neurons and their connections.} \label{fig:SingleHat}				
\end{figure}
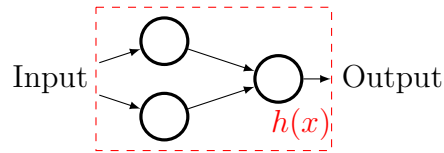

By comparing this with Equation \ref{eq:singleneuron}, we can see that this is equivalent to connecting three neurons as in Figure \ref{fig:SingleHat} and setting the weights and biases to the correct values by hand. We often refer to a group of neurons as a layer, where those that connect to the input are the input (first) layer, the neurons that connect to the input layer are the second layer and so forth, up until the last one that produces an output, also known as the output layer. Every layer that is neither an input or an output layer is also referred to as a hidden layer.

In essence, this is our first neural network that takes some value $x$ as input and returns $1$ if it is exactly $c_i$ or something less than $1$ or even $0$ if it is not (we can see this by plugging in values by hand or taking a look back at Figure \ref{fig:HatFunction}). Essentially, we made an $c_i$ detector, as that is the only value that returns $1$. Again, not very exciting. 

\begin{figure}[H]
	\centering
		\begin{tikzpicture}
		\draw[very thin,color=gray!60] (-6.1,2.0-0.1) grid (-6.0+3.2,2.0+2.2);
		\draw[very thick,->] (-6.4,2.0) -- (-6.0+3.6,2.0+0) node[right] {$x$};
		\draw[very thick,->] (-6,2.0-0.2) -- (-6.0+0,2.0+2.6) node[above] {$h_i(x)$};
		\draw[color=red] plot[domain=-6:-3,id=fx3] function{x<=-5 ? (x+6) + 2.05 : x<=-4 ?  -(x+4) + 2.05 : 2.05}; 
		
		\draw[very thin,color=gray!60] (-6.1,-2.0-0.1) grid (-6.0+3.2,-2.0+2.2);
		\draw[very thick,->] (-6.0,-2.0) -- (-6.0+3.6,-2.0+0) node[right] {$x$};
		\draw[very thick,->] (-6,-2.0-0.2) -- (-6.0+0,-2.0+2.6) node[above] {$h_{i+1}(x)$};
		\draw[color=red] plot[domain=-6:-3,id=fx4] function{x<=-5 ? -1.95 : x<=-4 ? (x+5)  - 1.95 : -(x+3) - 1.95}; 
		
		\draw[very thin,color=gray!60] (1.0-0.1,-0.1) grid (4.2,3.2);
		\draw[very thick,->] (1.0-0.2,0) -- (4.6,0) node[right] {$x$};
		\draw[very thick,->] (1.0,-0.2) -- (1.0,3.6) node[above] {$f(x)$};
		\draw[color=blue] plot[domain=1:4,id=sin1] function{sin((x-1))*3}; 
		\draw[color=red] plot[domain=1:4,id=fx5] function{x<=2 ? sin(1)*(x-1)*3 : x<=3 ? (sin(2)-sin(1))*(x-2)*3 + sin(1)*3 : (sin(3)-sin(2))*(x-3)*3 + sin(2)*3}; 
		\draw (2,-0.5) node {$a_i$};
		\draw (3,-0.5) node {$a_{i+1}$};
		
		\draw[very thick,->] (-2.0,3.0) -- (0.5,2.5);
		\draw[very thick,->] (-2.0,-1.0) -- (0.5,-0.5);
		\draw (-0.75,3.2) node {$\cdot g(a_i)$};
		\draw (-0.75,-0.3) node {$\cdot g(a_{i+1})$};
		\draw (-1,1) node {$+$};
		\end{tikzpicture}
	\caption{Piecewise approximation of some arbitrary function (blue) using multiple hat functions (red).} \label{fig:HatFunctions}
\end{figure}
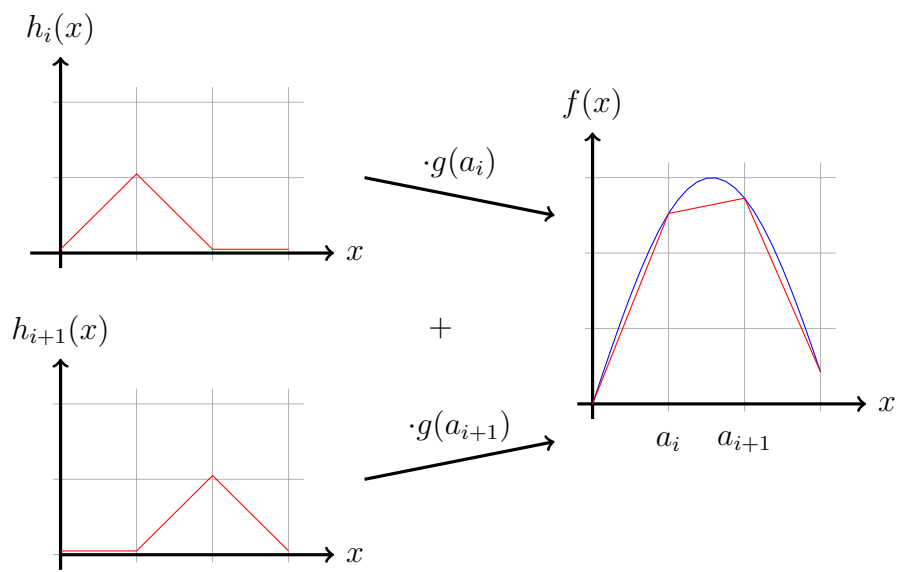

The beauty of hat functions, however, is that we can simply add up multiple copies of them create a piecewise linear approximation of some other function $g(x)$, as shown in Figure \ref{fig:HatFunctions}. All that needs to be done is to choose $g(c_1), \cdots, g(c_n)$ as the height of each hat function peak.

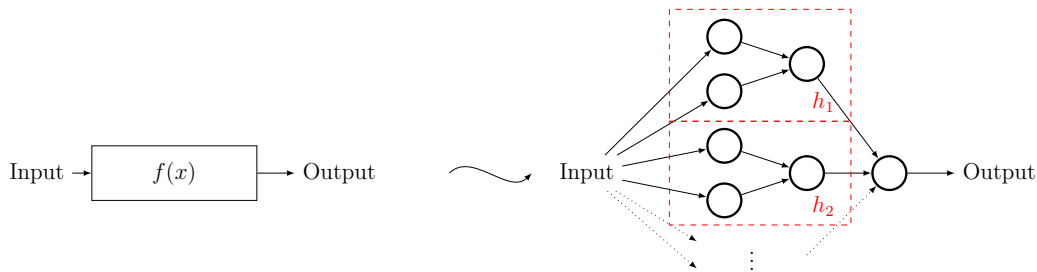
\begin{figure}[H]
	\centering
	\begin{adjustbox}{max width=\textwidth}
		\begin{tikzpicture}
		
		\node at (-5, 0)  {Output}  
		child[grow=left]
		{ 
			node [block] at (-1.5, 0) {$f(x)$} edge from parent[toparent] 
			child[grow=left]{ node (C) at (-1, 0) {Input} edge from parent[toparent] }
		};
		
		\node (F) at (7, 0)  {Output}  
		child[grow=left]
		{
			node (E) [whitenode] at (-0.5, 0)  {} edge from parent[toparent]
			child[grow=left]
			{ 
				node [whitenode] at (0, 2.0) {} edge from parent[toparent]
				child[grow=left]{ node (D) [whitenode] at (0, 0.5) {} edge from parent[toparent] }
				child[grow=left]{ node (C) [whitenode] at (0, -0.5) {} edge from parent[toparent] }
			}
			child[grow=left]
			{ 
				node [whitenode] at (0, 0.0) {} edge from parent[toparent]
				child[grow=left]
				{ 
					node [whitenode] at (0, 0.5) {} edge from parent[toparent] 
					child[grow=left]
					{
						node (A) at (-1, -0.5) {Input} edge from parent[toparent]
					}
				}
				child[grow=left]{ node (B) [whitenode] at (0, -0.5) {} edge from parent[toparent] }
			}
		};
		
		\draw[tochild, black] (A) -- (B);
		\draw[tochild, black] (A) -- (C);
		\draw[tochild, black] (A) -- (D);
		\draw[dotted, tochild, black] (A) -- (1.5, -1.25);
		\draw[dotted, tochild, black] (A) -- (1.5, -1.75);
		\draw[dotted, tochild, black] (3.5, -1.5) -- (E);
		
		\draw[tochild, black] (-3.0, 0.0) to [out=45,in=225] (-1.5, 0.0);
		
		\node [black] at (2.5, -1.5) {$\vdots $};
		
		\draw [dashed, red] (1,0.95) -- (4.3,0.95) -- (4.3,3) -- (1,3) -- (1,0.95);
		\node [red] at (3.8,1.3) {$h_1$};
		
		\draw [dashed, red] (1,0.95) -- (4.3,0.95) -- (4.3,-0.95) -- (1,-0.95) -- (1,0.95);
		\node [red] at (3.8,-0.6) {$h_2$};

		\end{tikzpicture}
	\end{adjustbox}
	\caption{Approximation of a non-linear block (left) by a network of classical neurons with ReLU activation functions (right). The individual hat functions are highlighted as red dashed rectangles and aggregated using an additional neuron.} \label{fig:CapsuleToHatNetwork}				
\end{figure}

This is equivalent to having the neural network shown in Figure \ref{fig:CapsuleToHatNetwork}, with all the weights and biases set to the appropriate values. Again, this is all still done by hand, which is not what we actually want.

\begin{figure}[H]
	\centering
		\begin{tikzpicture}
		
		\node (E) at (5.5, 0)  {Output 1}  
		child[grow=left]
		{ 
			node [whitenode] at (-0.5, 0.0) {} edge from parent[toparent]
			child[grow=left]
			{ 
				node (C) [whitenode] at (0, 0.0) {} edge from parent[toparent] 
				child[grow=left]
				{
					node (A) at (-0.5, -0.5) {Input} edge from parent[toparent]
				}
			}
			child[grow=left]
			{ 
				node (B) [whitenode] at (0, -1) {} edge from parent[toparent] 
			}
		};
		
		\node at (5.5, -1)  {Output 2}
		child[grow=left]
		{
			node (G) [whitenode] at (-0.5, 0.0) {} edge from parent[toparent]
		};
		
		\draw[tochild, black] (A) -- (B);
		\draw[tochild, black] (C) -- (G);
		\draw[tochild, black] (B) -- (G);

		\node [black] at (2.5, 1) {$\vdots $};
		\node [black] at (2.5, -2) {$\vdots $};

		\draw [dashed, red] (1,0.5) -- (4.3,0.5) -- (4.3,-0.45) -- (2.7,-0.45) -- (2.7,-1.5) -- (1,-1.5) -- (1,0.5);	
		\draw [dotted, blue] (1.05,0.45) -- (2.65,0.45) -- (2.65,-0.55) --(4.3,-0.55) -- (4.3,-1.45) -- (1.05,-1.45) -- (1.05,0.45);
		\node [red] at (3.8,0.8) {$h_i$};
		\node [blue] at (3.8,-1.9) {$\tilde{h}_i$};
		
		\end{tikzpicture}
	\caption{Example of a single hat function (red, dashed) being reused (blue, dotted) to produce a different output.} \label{fig:MultipleOutputs}				
\end{figure}
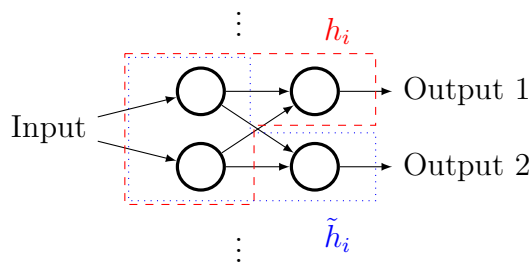

This idea can be easily extended to higher-dimensional hat functions with multiple inputs and multiple outputs (As an example, see the case of two outputs in Figure \ref{fig:MultipleOutputs}). Doing this allows us to construct a neural network that can approximate any function. As a matter of fact, the more neurons we add to this network, the closer we can get to the function we want to approximate. In essence we have explored how neural networks can be universal function approximators \cite{Csaji:2001}. 

But everything discussed so far is \textbf{not} how neural networks are designed and constructed. In reality, neurons are connected and then trained, without knowing the actual function $g(\vec{x})$ we've used thus far. The training process essentially takes a bunch of data and attempts to find the best weights and biases to approximate this data by itself, replacing our hand-designed approach. In general, these weights and biases will not resemble hat functions or anything similar. But, as we are still using ReLU, the final function described by a network that was trained instead of hand-designed, will still look like a piecewise linear interpolation. It just happens that the training process automatically found optimal support points (similar to those of our hat functions) based on the training data.

Let's describe this training process in more detail. It relies on a mathematical tool called backpropagation \cite{Goodfellow:2016}. Imagine neural networks and backpropagation as an assembly line of untrained workers that want to build a smartphone. The last employee (layer) of this assembly line only knows what the output should be, but not how to get there. He looks at a finished smartphone (output) and deduces that he needs some screen and some "back" element that holds the screen. So, he turns to the employee that is just to his left and tells him: "You need to produce a screen and a back element". This employee, knowing nothing, says "sure", looks at these two things and tries to break it down even further. He turns to his left to tell the next employee, that he needs "a piece of glass, a shiny metal and some rubber" and that neighbor will say "sure". This goes on all the way through the assembly line to the last employee, who is totally confused how he should get his hands on "a diamond, a car and some bronze swords", if all he has is copper, silicon and glass (inputs). He won't say "sure", as he can't make what his neighbor wants. So he tells him what he can make. At this point, the foreman will step in and tell them to start the process over, but to keep in mind what they learned the first time around. Over time, the assembly line employees will slowly figure a way out that works. 

So how does backpropagation really work? The idea is to have a measure of how well the current model approximates the "true" data. This measure is called the loss function. Assume we have a training pair of inputs $\vec{x}$ and the corresponding correct outputs $\vec{y}$. This can be an image ($\vec{x}$) and what type of object it is ($\vec{y}$). Now, when we input $\vec{x}$ into our neural network model, we get some output $\vec{\tilde{y}}$, which is most likely very different to the correct value $\vec{y}$, if we haven't trained it yet. The loss function $l$ assigns a value to the difference between the true $\vec{y}$ and the one our model calculates at this exact moment $\vec{\tilde{y}}$. How we define this loss is up to us and will have different effects on training later on. A simple example for a loss function is the square loss 
\begin{equation}
l(\vec{y}, \vec{\tilde{y}}) = (\vec{y} - \vec{\tilde{y}})^2 \;\;\; .
\end{equation}

It makes sense to rewrite this in a slightly different way. Let's use $f$ to denote our neural network model and $\vec{\theta} = [w_0, \cdots, w_n, b_0, \cdots, b_m]$ to be a vector of all our weights and biases. We write $\vec{\tilde{y}} = f(\vec{x} \vert \vec{\theta})$ to mean that our model produces the output $\vec{\tilde{y}}$ from inputs $\vec{x}$ based on the weights and biases $\vec{\theta}$. In our example, this is equivalent to the foreman complaining: "If you continue with your work in this way ($\vec{\theta}$), the smartphones you produce ($\vec{\tilde{y}}$)  from this set of raw materials ($\vec{x}$) will look only 25\% ($l$) like the smartphone we are meant to produce ($\vec{y}$)."

We, however, have a lot of data points $(\vec{x}_i, \vec{y}_i)$ and want some quantity that measures how the neural network performs on these as a whole, where some might fit better than others. This is called the cost function $C(\vec{\theta})$, which again depends on the model parameters. The most obvious would be to simply add up all the square losses of the individual data points and take the mean value. As a matter of fact, that is exactly what happens with the mean squared error (MSE):
\begin{equation}\label{eq:MSE}
\text{MSE}(\vec{\theta}) = \frac{1}{n} \sum_i^n (\vec{y}_i - f(\vec{x}_i \vert \vec{\theta}))^2 = \frac{1}{n} \sum_i^n l(\vec{y}, f(\vec{x}_i \vert \vec{\theta})) \;\;\; .
\end{equation}

As the name suggests, the cost measures the mean of all the individual losses. In our example, this is equivalent to the foreman calculating how bad the employees have performed over an entire batch of smartphones. Note that it is quite typical to write the cost function name $\text{MSE}$ instead of $C$. Further, because they are so similar in nature, in a lot of articles the words "loss" and "cost" are used interchangeably and it becomes clear from context which is meant.

\begin{figure}[H]
	\centering
		\begin{tikzpicture}
		
	\draw[very thick,->] (-0.2,0) -- (10,0) node[right] {$w$};
	\draw[very thick,->] (0.0,-0.2) -- (0.0,4.0) node[above] {$C(w)$};
	\draw[color=blue] plot[domain=0:10,id=sin2] function{-sin(x) + x*0.3 + 1};

	\draw[color=red, dashed, very thick] (1.3,3.0) -- (1.3,-0.2) node[below] {$\tilde{w}_a$}; 
	\draw[color=red, dashed, very thick] (7.6,3.0) -- (7.6,-0.2) node[below] {$\tilde{w}_b$}; 
		
		\end{tikzpicture}
	\caption{An example cost function $C$ for the simplest case of a single weight parameter $w$. We find at least two local minima $\tilde{w}_a$ and $\tilde{w}_b$, where $\tilde{w}_a$ might even be a global minimum.} \label{fig:SingleVarCost}				
\end{figure}
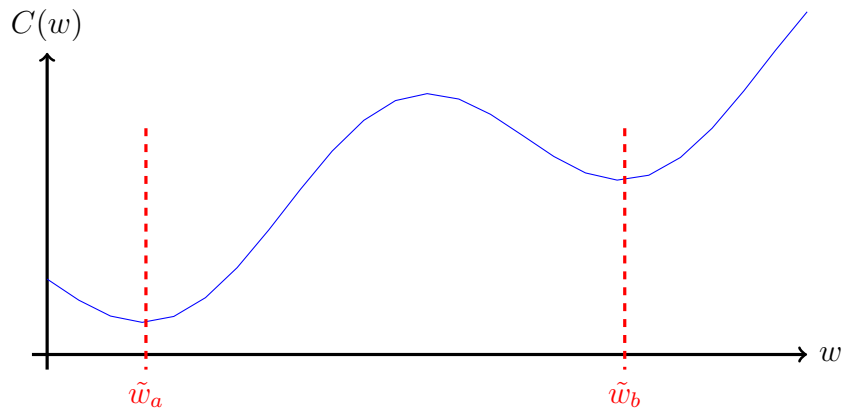

Again, this cost function measures how far off we are with our model, based on the current weights and biases $\vec{\theta}$ we are using. Ideally, if we have a cost of $0$, that would mean we are as close as possible with our model to the training data. This, however, is seldom possible. Instead, we will settle for parameters $\vec{\theta}$ that minimize the cost as much as possible and our goal is to change our weights and biases in such a way, that the value of the cost function goes down. Let's take a look at the simplest case and pretend that we only have a single parameter in $\vec{\theta}$. In this case, we can imagine the cost as a simple function over a single variable, say one single weight $w$, as shown in Figure \ref{fig:SingleVarCost}.

From the graph we see that we achieve the lowest cost if our weight has a value of $\tilde{w}_a$. However, in reality, we don't see this graph. In order to draw this graph, we had to compute the cost for every possible value of $w$. This is fine for a single variable, but our neural network might have millions of weights and biases. There isn't enough compute power in the world to try out every single value combination for millions of such parameters. 

\begin{figure}[H]
	\centering
		\begin{tikzpicture}
		
		\draw[very thick,->] (-0.2,0) -- (10,0) node[right] {$w$};
		\draw[very thick,->] (0.0,-0.2) -- (0.0,4.0) node[above] {$C(w)$};
		\draw[color=blue] plot[domain=0:10,id=sin2] function{-sin(x) + x*0.3 + 1};

		\draw[color=red, dashed, very thick] (3.9,3.0) -- (3.9,-0.2) node[below] {Start}; 
		%\draw[color=red, dashed, very thick] (3.9,3.0) -- (3.9,-0.2) node[below] {Start}; 
		
		\draw[color=red, very thick,->] (3.8,2.9) -- (3.4,2.4);
		\draw[color=red,very thick,->] (3.3,2.3) -- (2.9,1.8);
		\draw[color=red,very thick,->] (2.8,1.7) -- (2.4,1.2);
		
		\end{tikzpicture}
	\caption{Instead of looking for the minimum by checking all values individually, we begin at some point (start) and move "downhill" (red arrows) until we reach a minimum.} \label{fig:GradientDescent}				
\end{figure}
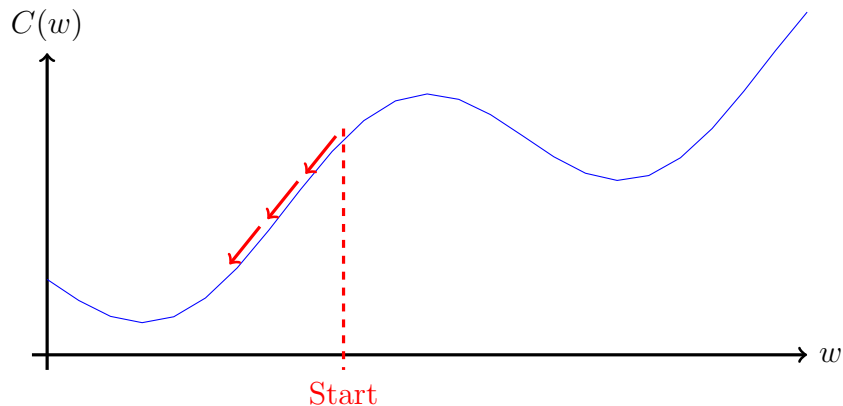

There is another solution to our problem. Consider Figure \ref{fig:GradientDescent}. 

Let's pretend we started with some value of $w$ that isn't at the minimal cost and we have no idea where it might be. What we do know is that the minimal cost is at the lowest point. If we can measure the slope at the point we are at at the moment, then at least we know in what direction we need to move in order to reduce our cost. And if we keep repeating this process, \ie, measuring the slope and going "downhill", we should reach the lowest point at some stage. 

This should be familiar from calculus. Measuring the slope in respect to some variable is nothing more than taking the derivative. In our case, that is $\frac{\partial C(w)}{\partial w}$. Let's rewrite this in terms of all parameters $\vec{\theta}$ and introduce the gradient
\begin{equation}
\nabla C(\vec{\theta}) = \left[\frac{\partial C(\vec{\theta})}{\partial w_0}, \cdots, \frac{\partial C(\vec{\theta})}{\partial w_n}, \frac{\partial C(\vec{\theta})}{\partial b_0}, \cdots, \frac{\partial C(\vec{\theta})}{\partial b_m}  \right] \;\;\; .
\end{equation}

To find better parameters that reduce the cost function, we simply move a tiny step of size $\alpha$ into the direction with the steepest gradient (going downhill the fastest). The equation for which looks like this:
\begin{equation}
\vec{\theta}^{\;\text{(new)}} = \vec{\theta}^{\;\text{(old)}} - \alpha \cdot \nabla C(\vec{\theta}^{\;\text{(old)}}) \;\;\; .
\end{equation}

This is called gradient descent, which is an optimization method. There are hundreds of variations of this method, which may vary the step size $\alpha$ (also known as the learning rate) or do other fancy things. Each of these optimizers has a name (Adam \cite{Kingma:2014}, RMSProp \cite{Hinton:2012}, etc.) and can be found in most deep learning libraries, but the exact details aren't important for this article.

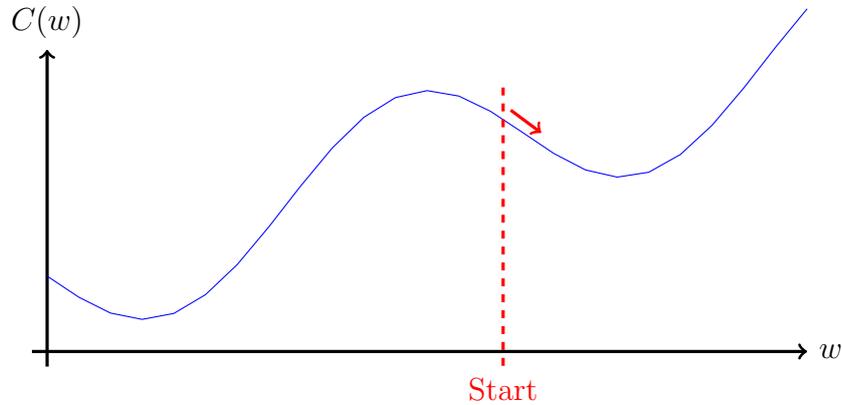
\begin{figure}[H]
	\centering
		\begin{tikzpicture}
		
		\draw[very thick,->] (-0.2,0) -- (10,0) node[right] {$w$};
		\draw[very thick,->] (0.0,-0.2) -- (0.0,4.0) node[above] {$C(w)$};
		\draw[color=blue] plot[domain=0:10,id=sin2] function{-sin(x) + x*0.3 + 1};

		\draw[color=red, dashed, very thick] (6.0,3.5) -- (6.0,-0.2) node[below] {Start}; 
		%\draw[color=red, dashed, very thick] (3.9,3.0) -- (3.9,-0.2) node[below] {Start}; 
		
		\draw[color=red, very thick,->] (6.1,3.2) -- (6.5,2.9);
		
		\end{tikzpicture}
	\caption{We started at a point that leads us to a local minimum. We might get stuck there and never even know that it is just a local minimum and not a global one.} \label{fig:LocalMini}				
\end{figure}

Using gradient descent, however, does not guarantee we find the lowest possible cost (also known as the global minimum). Instead, if we take a look at Figure \ref{fig:LocalMini}, we see that we might get stuck in some valley that isn't the lowest minimum, but only a local minimum. Finding the global minimum is probably the largest unsolved challenge in machine learning and can play a role in the upcoming exercises.

Next, all we need to do is to calculate $\nabla C(\vec{\theta})$, which is done iteratively by backpropagation. We won't go into the exact detail of the algorithm, as it is quite technical, but rather try to give some intuitive understanding of what is happening. At first, finding the derivative of $C$ seems daunting, as it entails finding the derivative of the entire neural network. Let's assume we have a single input $x$, a single output $y$ and a single neuron with activation function $a$ and weight $w$ with no bias, \ie, the example we have been working with so far. We want to find $\frac{\partial C(w)}{\partial w}$, which seems hard. But, notice that $C$ contains our model $f = a(w \cdot x)$, which means we can apply the chain rule in respect to $a$:
\begin{equation}
\frac{\partial C(w)}{\partial w} = \frac{\partial C(w)}{\partial a}\frac{\partial a}{\partial w} \;\;\; .
\end{equation}
This is quite a bit easier to solve. As a matter of fact, if we are using the MSE cost, we quickly find the derivative of Equation \ref{eq:MSE} to be
\begin{equation}
\frac{\partial C(w)}{\partial a} = (y - a) \;\;\; .
\end{equation}
Now we just need to find $\frac{\partial a}{\partial w}$. Recall that we introduced $z = \vec{w}\cdot\vec{x} + b$ as an intermediary step in a neuron (Equation \ref{eq:ZZ}), which is the value just before it is piped through an activation function $a$. For our case we just have $z = wx$. We use the chain rule again, but this time in respect to $z$:
\begin{equation}
\frac{\partial a}{\partial w} = \frac{\partial a}{\partial z}\frac{\partial z}{\partial w} \;\;\; .
\end{equation}
We find that $\frac{\partial a}{\partial z}$ is just the derivative of the ReLU activation function, which we can look up:
\begin{equation}
\frac{\partial a}{\partial z}=\begin{cases}
0 \;\;\; \text{for} \; x\leq 0  \\
1 \;\;\; \text{for} \; x>0
\end{cases}
\;\;\; 
\end{equation}
and $\frac{\partial z}{\partial w} = x$ (we ignore the "minor" detail that obviously ReLU isn't differentiable at $0$...). Multiplying it all together and we have found our gradient!

Now, what happens when we add more layers to the neural network? Well, our derivative $\frac{\partial z}{\partial w}$ will no longer just be $x$, but rather, it will be the activation function of the lower layer, \ie, $\frac{\partial z}{\partial w} = a^{\text{(lower layer)}}(\cdots)$! From there, we basically start back at $\frac{\partial a}{\partial w}$, just with the values of the lower layer. This in essence is the reason why this algorithm is called backpropagation. We start at the last layer and, in a sense, optimize beginning from the back. Now, apart from going deeper, we can also add more weights to each neuron and more neurons to each layer. This doesn't change anything really. We only need to keep track of more and more indices and we have our backpropagation algorithm.

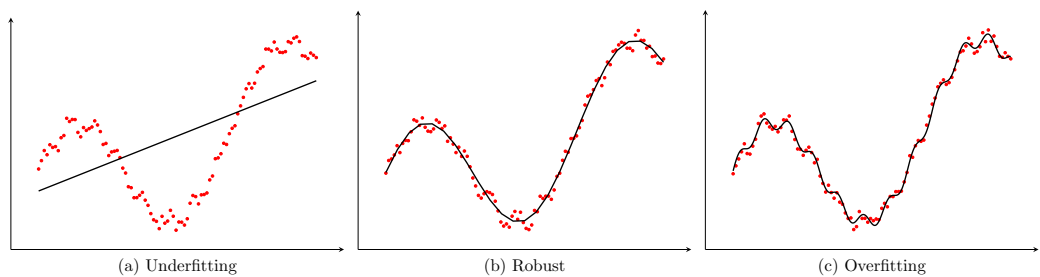
\begin{figure}[H]
	\centering
	\begin{adjustbox}{max width=\textwidth}
		\subfloat[Underfitting]{
			\begin{tikzpicture}[
			declare function={a(\x)=-cos(120*x) + x*0.4 + 0.5;},
			declare function={b(\x)=-cos(120*x) + cos(1000*x)*0.1 + x*0.4 + 0.5;},
			declare function={c(\x)= x*0.4 + 0.75;},
			baseline=(current axis.south)%,scale=0.3
			]
			\begin{axis}[
			domain=1:5,
			axis lines=middle,
			axis equal image,
			xtick=\empty, ytick=\empty,
			enlargelimits=true,
			clip mode=individual, clip=false
			]
			\addplot [red, only marks, mark=*, samples=100, mark size=0.75] {b(x) + 0.2 * (0.5 * rand) };
			%\addplot [thick] {a(x)};
			%\addplot [thick, samples=500] {b(x)};
			\addplot [thick] {c(x)};
			\end{axis}
			\end{tikzpicture}
		}
		\subfloat[Robust]{
			\begin{tikzpicture}[
			declare function={a(\x)=-cos(120*x) + x*0.4 + 0.5;},
			declare function={b(\x)=-cos(120*x) + cos(1000*x)*0.1 + x*0.4 + 0.5;},
			declare function={c(\x)= x*0.4 + 0.75;},
			baseline=(current axis.south)%,scale=0.3
			]
			\begin{axis}[
			domain=1:5,
			axis lines=middle,
			axis equal image,
			xtick=\empty, ytick=\empty,
			enlargelimits=true,
			clip mode=individual, clip=false
			]
			\addplot [red, only marks, mark=*, samples=100, mark size=0.75] {b(x) + 0.2 * (0.5 * rand) };
			\addplot [thick] {a(x)};
			%\addplot [thick, samples=500] {b(x)};
			%\addplot [thick] {c(x)};
			\end{axis}
			\end{tikzpicture}
		}
		\subfloat[Overfitting]{
			\begin{tikzpicture}[
			declare function={a(\x)=-cos(120*x) + x*0.4 + 0.5;},
			declare function={b(\x)=-cos(120*x) + cos(1000*x)*0.1 + x*0.4 + 0.5;},
			declare function={c(\x)= x*0.4 + 0.75;},
			baseline=(current axis.south)%,scale=0.3
			]
			\begin{axis}[
			domain=1:5,
			axis lines=middle,
			axis equal image,
			xtick=\empty, ytick=\empty,
			enlargelimits=true,
			clip mode=individual, clip=false
			]
			\addplot [red, only marks, mark=*, samples=100, mark size=0.75] {b(x) + 0.2 * (0.5 * rand) };
			%\addplot [thick] {a(x)};
			\addplot [thick, samples=500] {b(x)};
			%\addplot [thick] {c(x)};
			\end{axis}
			\end{tikzpicture}
		}
	\end{adjustbox}
	\caption{Piecewise approximation of some arbitrary function (blue) using multiple hat functions (red).} \label{fig:OverUnderFitting}
\end{figure}

When designing a neural network and deciding upon how long and with what parameters to run the optimizer, we need to keep in mind the concept of under- and overfitting. Figure \ref{fig:OverUnderFitting} should give some intuition what both these terms mean. 

If, for example, our model just doesn't have enough neurons to approximate the function, the optimizer will try its best to come up with something that is close enough, but just doesn't represent the data well enough (underfitting). As another example, if we have a model that has more than enough parameters available and we train it on data far too long, the optimizer will get too good at fitting to the data, so much so, that the neural network almost acts like a look-up table (overfitting). Ideally, we want something in between under- and overfitting. However, in later sections we will see, that for our purposes, it isn't necessarily bad to overfit.

This wraps up the very basics of neural networks we are going to cover and we move one level higher to see what we can do with them from a network architecture point-of-view. Roughly speaking, we can perform two different tasks with a network, regression and classification:

\begin{itemize}
	\item \textbf{Regression} allows us to uncover the relation between input variables and do predictions and interpolations based on them. A classical example would be predicting stock prices. Generally, the result of a regression analysis is some vector of continuous values.

	\item \textbf{Classification} allows us to classify what category a set of inputs might belong to. Here, a typical example is the classification of what object is in an image. The result of this analysis is a vector of probabilities in the range of $0$ and $1$, ideally the sum of which is exactly $1$ (also referred to as a multinomial distribution).
\end{itemize}

Thus far, our introduction of neural networks has covered how regression works. To do classification, we need just a few more ingredients. Obviously, we want our output to be a vector of mainly $0$s and $1$s. Using ReLU on the hidden layers is fine, but problematic on the last layer, as it isn't bounded and doesn't guarantee that the output vector sums to $1$. For this purpose, we have the softmax function
\begin{equation}\label{eq:Softmax}
\text{softmax}(z_0, \cdots, z_j)_i = \frac{e^{z_i}}{\sum_j^n e^{z_j}} \;\;\; .
\end{equation}
It is quite different to all other activation functions, as it depends on \textbf{all} $z$ values from all the output neurons, not just its own. But it has to, as otherwise it can't normalize the output vector to add up to be exactly $1$.

Even with softmax in place, training this model would be problematic. The loss function we have used thus far, MSE, is ill suited for this task, as miss-classifications aren't penalized enough. We need some loss that takes into account that we are comparing a multinomial distribution. Luckily, there exists a loss function that does exactly that, namely the cross-entropy loss
\begin{equation}
l(\vec{y}, f(\vec{x}_i \vert \vec{\theta})) = - \vec{y} \cdot \log(f(\vec{x}_i \vert \vec{\theta})) \;\;\; .
\end{equation}
The exact details why the loss is the way it is, isn't important for our purposes. We recommend Aurélien Géron's video \cite{Geron:2018} on the subject as an easy to understand explanation for the interested reader.

With the basic differences between classification and regression covered, we move on to the different types of layers found in a neural network. So far we have covered what is called a dense layer, \ie, every neuron of one layer is connected to (almost) every neuron of the next layer. These are perfectly fine and in theory, we are able to do everything with just these types of layers. However, having a deep network of nothing more than dense layers greatly decreases the performance while training and when using it later on. Instead, using some intuition and reasoning, a lot of new types of layers were introduced that reduced the number of connections to a fraction. In our motivation using hat functions, we saw that it is possible to do quite a lot without connecting every neuron with every other neuron.

Let's take the example of image classification. Images are comprised of a lot of pixels. Even if we just have a small image of size $28\times 28$, we are already looking at $784$ pixels. For a dense layer and just a single neuron for each pixel, we are already looking at $614656$ connections, \ie, $614656$ different weights, for a single layer. But we can be a bit smarter about this. We can make the assumption, that each pixel has to be seen in context to the neighboring pixels, but not those far away. So, instead of connecting each of the neurons to every pixel, we just connect the neurons to one pixel and the $8$ surrounding ones. We still have $784$ neurons, but reduced the amount of weights to $784\times 8 = 7056$, as each neuron is only connected to $9$ pixels in total.

\begin{figure}[H]
	\centering
	\includegraphics[width=1.0\textwidth]{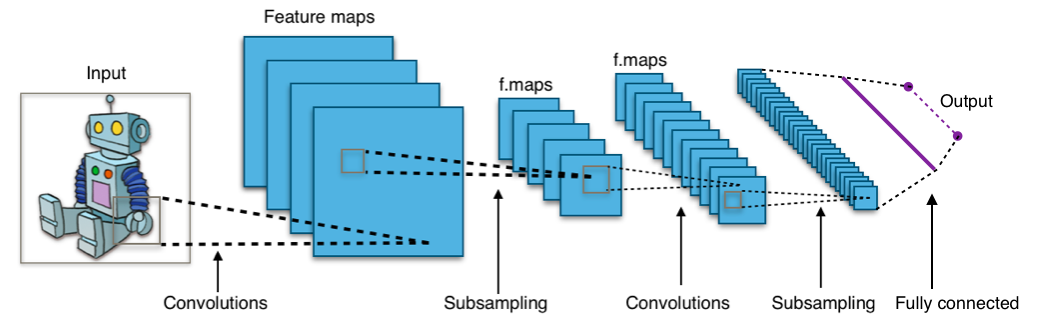}
	\caption{A typical convolutional neural network. Image by Aphex34 (\url{https://en.wikipedia.org/}). } \label{fig:ConvNet}
\end{figure}

In some sense, we have made $28\times 28$ small $3\times 3$ regions inside of the image. That's still a lot. We can aggregate these results and subsample by a factor of $2$, so that the following layer only needs $14 \times 14$ neurons. Now, we repeat this process and connect these $14 \times 14$ to $3 \times 3$ of the neurons of the previous layer. In essence, we are relating the small $3\times 3$ regions of the input image to other such regions (See Figure \ref{fig:ConvNet} for an illustration). We then subsample some more and are left with a grid of $7 \times 7$ neurons (depending on the padding we used). With so few neurons left, it is computationally not too expensive to start adding dense layers from this point onwards.

What we described here are called convolutional layers \cite{LeCun:1989} and the small regions they process are their filters and the result of each filter is referred to as a feature. In our construction we only added one neuron for each filter, but it is perfectly fine to add more. This entire process, including subsampling can be seen in Figure \ref{fig:ConvNet}. What we described as subsampling is also often referred to as pooling or down-sampling. There are different methods to perform subsampling, such as max-pooling \cite{Yamaguchi:1990}, which just takes the largest value in some grid.

Convolutional layers are de-facto standard in image classification and have found their use in non-image related tasks, such as text classification. The development of new architectures using convolutional layers is too rapid to name them all. There have, however, been a lot of milestones worth mentioning and looking into, such as LeNet \cite{LeCun:1998}, AlexNet \cite{Krizhevsky:2012}, GoogleNet \cite{Szegedy:2015} and ResNet \cite{He:2016}.

\begin{figure}[H]
	\centering
	\includegraphics[width=1.0\textwidth]{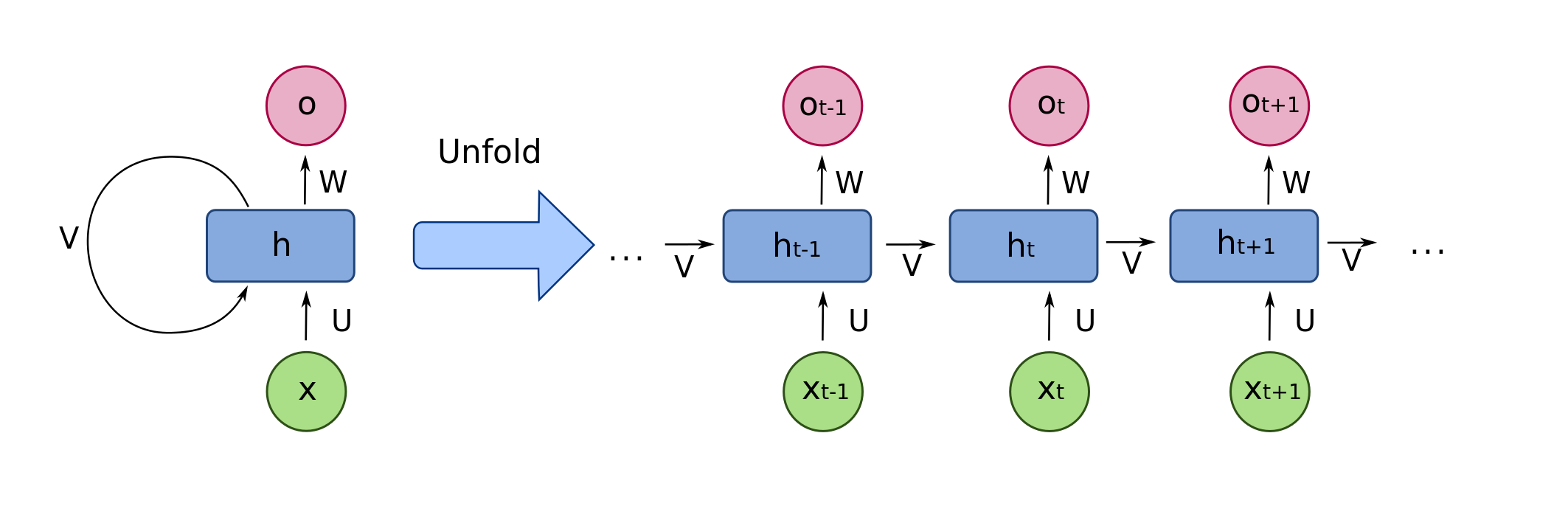}
	\caption{The general structure of recurrent neural networks. Image by François Deloche (\url{https://en.wikipedia.org/}). } \label{fig:RNN}
\end{figure}

So far, we have looked at purely feed-forward networks, which take an input, process it and produce an output. However, we want to mention what happens when we have a network that takes some input, processes it and produces an output, but then continues taking inputs for processing and always "remembers" what it did at the last step. This remembering is equivalent to passing on some hidden state to the next step. A schematic illustration of such a network is found in Figure \ref{fig:RNN}. This type of network is called a recurrent neural network (RNN). By passing on a hidden state, this network is ideal for sequence data. As such, they are commonplace in natural language processing (NLP), where the network begins by processing a word or sentence and passes on the information to the next part of the network when it looks at the following word or sentence. After all, we do not forget the beginning of a sentence while reading it.

One of the first and most famous type of RNN is the long-short term memory (LSTM) \cite{Hochreiter:1997}, which is the basis for a lot of newer models, such as Seq2Seq \cite{Sutskever:2014} for translating/converting sequences of text into other sequences. 

\subsection{How it works}

Before we begin to dive into some attack methods, here is a quick breakdown of a couple of the more interesting security implementations that employ neural networks. There are, of course, many more applications apart from those listed in the following. Network scanners, web application firewalls and more can all be implemented with at least some part using deep learning. The ones we discuss here are interesting, as the methods and exercises presented in later sections are mainly aimed at toy versions of their actual real-life counter-part.

We won't go deep into each application, but rather discuss one specific implementation for each. For a review of deep learning methods found in cyber security, we refer to the survey by Berman \etal \cite{Berman:2019}. Isao Takaesu, the author of DeepExploit we highlight in this section, also has a more in-depth course on the defensive aspects of machine learning \cite{Takaesu2:2019}.

\subsubsection{Biometric Scanners}

There are a lot of different biometric scanners. We will be looking at iris scanners for security access (\ie, one that tells us "access" or "no access") based on deep learning. The naive approach to implementing such a scanner would be to train a CNN on a large set of irises for "access" and another for "no access".

This type of approach would have multiple problems, some of which will become clear through the exercises later on. From a practical standpoint alone, this would be unfeasible, as each new person who needs his iris to grant "access" would require the neural network to be re-trained.

\begin{figure}[H]
	\centering
	\includegraphics[width=1.0\textwidth]{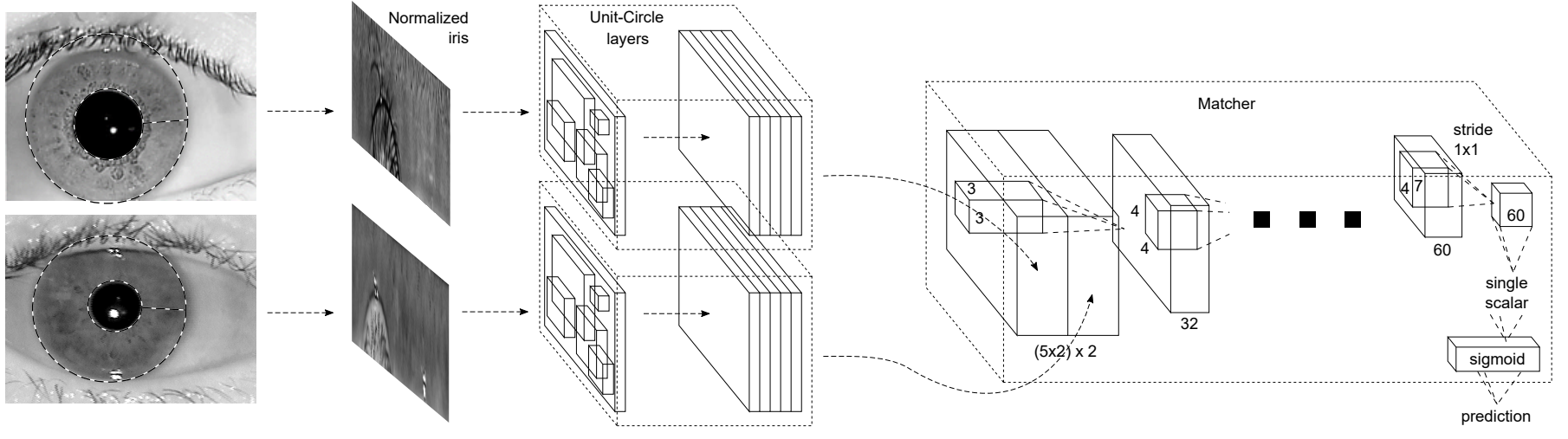}
	\caption{Iris verification with IrisMatch-CNN. Two irises are detected and normalized. The normalized irises are fed into the Unit-Circle (U-C) layers. The responses from the U-C layers are concatenated and fed into the Matcher convolutional network. A single scalar is produced – the probability of a match. Two irises match if the probability is greater than a given threshold. Figure and description from \cite{Spetlik:2019}.} \label{fig:Iris}
\end{figure}

A more sensible solution is presented by Spetlik and Razumenic \cite{Spetlik:2019}. An overview of the architecture is found in Figure \ref{fig:Iris}.

The description of Figure \ref{fig:Iris} should yield some familiar terms. In essence the iris is broken down into a set of features (we can think of them like, iris color, deformations, etc.) using their U-C layers, which are their custom version of convolutional layers. This is done for the input iris from the current scan and a reference iris stored in some database. The resulting features from both these irises are then both fed through another convolutional network, the matcher. As the name implies, this CNN is responsible for matching the input iris to the reference based on the features and produce a single output value: match (1) or no match (0).

The major advantage of this approach is, that it only needs to be trained once, as the network itself doesn't grant access, but rather checks if two irises match. Adding new irises, thus, only requires adding it to the reference database. 

However, it still needs to be trained to be able to extract features from an image of an iris and to perform matching. Luckily, there are available datasets for iris images available online, such as CASIA-IrisV4 \cite{Casia:2010}. 

\subsubsection{Intrusion Detection}

Most modern approaches to intrusion detection systems are indeed based on combinations of machine learning methods \cite{Garcia-Teodoro:2009}, such as support vector machines \cite{Li:2012}, nearest neighbors \cite{Lin:2015} and now deep learning. We will be focusing on the implementation by Shone \etal \cite{Shone:2018}.

Intrusion detection systems need to be fast to handle large volumes of network data, especially if they are meant for a real-time application instead of forensics. Any deep learning implementation should therefore be compact. However, it must still be able to handle the diversity of the data and protocols found in a network.

\begin{figure}[H]
	\centering
	\includegraphics[width=1.0\textwidth]{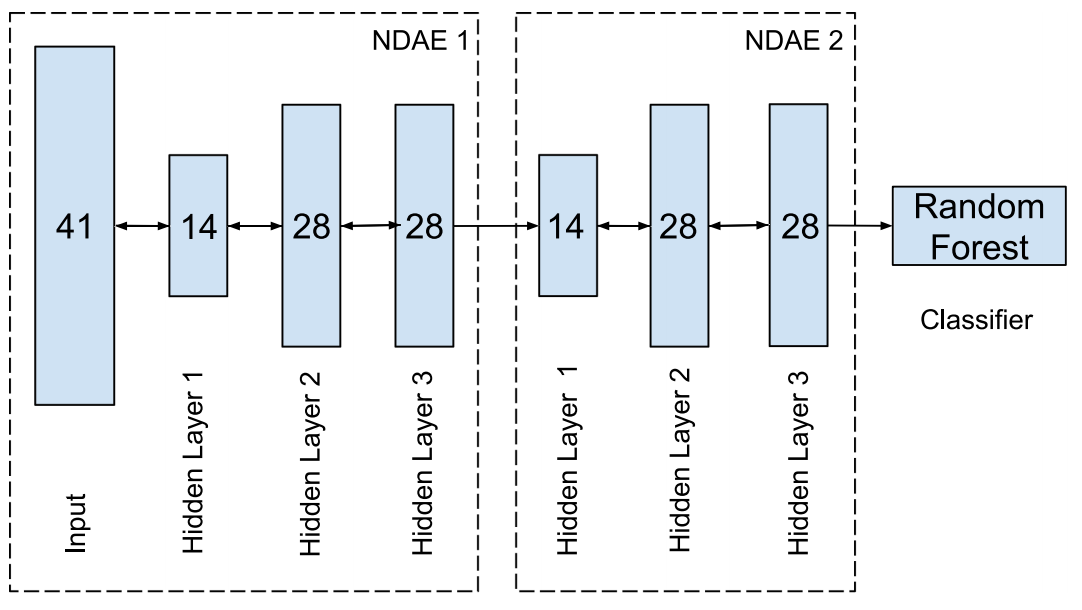}
	\caption{The proposed intrusion detection architecture. Figure from \cite{Shone:2018}.} \label{fig:Nids}
\end{figure}

The architecture proposed by Shone \etal shown in Figure \ref{fig:Nids} is compact enough to do fast computations. The overall idea is to use a neural network to take the 41 input features of the KDD1999/NSL-KDD dataset \cite{KDD:1999}\cite{Cybersecurity:2019} found in Table \ref{tab:Kdd} and encode them into a smaller set of 28 features, which are better suited for classification using a different machine learning method, random forest. In other words, the neural network is used basically to pre-processes the data.

\begin{table}[H]
	\begin{center}
	\begin{tabular}{ || c c | c c || }
		
1 & duration & 22 & is\_guest\_login \\
2 & protocol\_type & 23 & count \\
3 & service & 24 & srv\_count \\
4 & flag & 25 & serror\_rate \\
5 & src\_bytes & 26 & srv\_serror\_rate \\
6 & dst\_bytes & 27 & rerror\_rate \\
7 & land & 28 & srv\_rerror\_rate \\
8 & wrong\_fragment & 29 & same\_srv\_rate \\
9 & urgent & 30 & diff\_srv\_rate \\
10 & hot & 31 & srv\_diff\_host\_rate \\
11 & num\_failed\_logins & 32 & dst\_host\_count \\
12 & logged\_in & 33 & dst\_host\_srv\_count \\
13 & num\_compromised & 34 & dst\_host\_same\_srv\_rate \\
14 & root\_shell & 35 & dst\_host\_diff\_srv\_rate \\
15 & su\_attempted & 36 & dst\_host\_same\_src\_port\_rate \\
16 & num\_root & 37 & dst\_host\_srv\_diff\_host\_rate \\
17 & num\_file\_creations & 38 & dst\_host\_serror\_rate \\
18 & num\_shells & 39 & dst\_host\_srv\_serror\_rate \\
19 & num\_access\_files & 40 & dst\_host\_rerror\_rate \\
20 & num\_outbound\_cmds & 41 & dst\_host\_srv\_rerror\_rate\\
21 & is\_host\_login 

	\end{tabular}
\end{center}

	\caption{Features of the network data found in the KDD1999 dataset \cite{KDD:1999}.} \label{tab:Kdd}
\end{table}

This idea of encoding the input features into something that is easier to work with is common in deep learning and we saw that the iris scanner essentially did the same with its feature extractor. 

Let's focus on the datasets used for training. From the table of features for the KDD1999/NSL-KDD dataset, it should be clear that this is a very shallow inspection of network traffic, where the packet's content is largely ignored. From the architecture we know inspection happens on a per-packet basis. This allows us draw some first conclusions: It does not take into account the context the packet was send in (\textit{"is this well-formed but unusual user behavior?"}) and the timing (\textit{"is it a beacon?"}).

Our main takeaway here is, that without deeper knowledge of what the layers do and how random forest works, we are already able to formulate possible attack plans, just by looking at the architecture and the training data. In contrast, were we to find LSTM layers in the neural network's architecture, it would give some indication that the system might be analyzing sequences of packets, possibly making it context-sensitive and mitigating our planned attack path.

Newer datasets, such as CICIDS2017 \cite{Cybersecurity:2019}, use \texttt{pcap} files. These reflect real world scenarios more accurately. CICIDS2017 contains the network behavior data for 25 targets going about their daily business or facing an actual attack. These attacks include all sorts of possible attack vectors, such as a DoS or an SQL injection. While this dataset does not cover the entire MITRE ATT\&CK Matrix \cite{MITRE:2019}, it does provide enough hints at what neural network based NIDS are possibly looking for.

% \blueteam{Bots trained on real-life frequency data can seem very human-like in their communication with C2 servers. Pure rule-based detection has to be quite sophisticated to catch this behavior. Machine learning can be helpful in discriminating between traffic generated by a real human and that generated by a bot.}
% Even though AI is on the package doesnt mean it has all the benefits of other AIs (sequence vs. packet inspection)

\subsubsection{Anti-Virus}

Anti-Virus software is another prominent type of application that utilizes machine learning. As previously, we follow a specific implementation for our discussion. Here we use the deep learning architecture proposed in \cite{Pascanu:2015}.
 
The approach they chose was to identify malicious code by what API calls it makes and in what order. To apply deep learning to this method, the API calls the code makes need to be preprocessed into a representation a neural network can understand. For this, one can use a vector, where each element in the vector represents one type of API call. We set all the elements of that vector to 0, except for the element that represents the API call made at the moment, which we set to 1. This is also referred to as one hot encoding. 

\begin{figure}[H]
	\centering
	\includegraphics[width=1.0\textwidth]{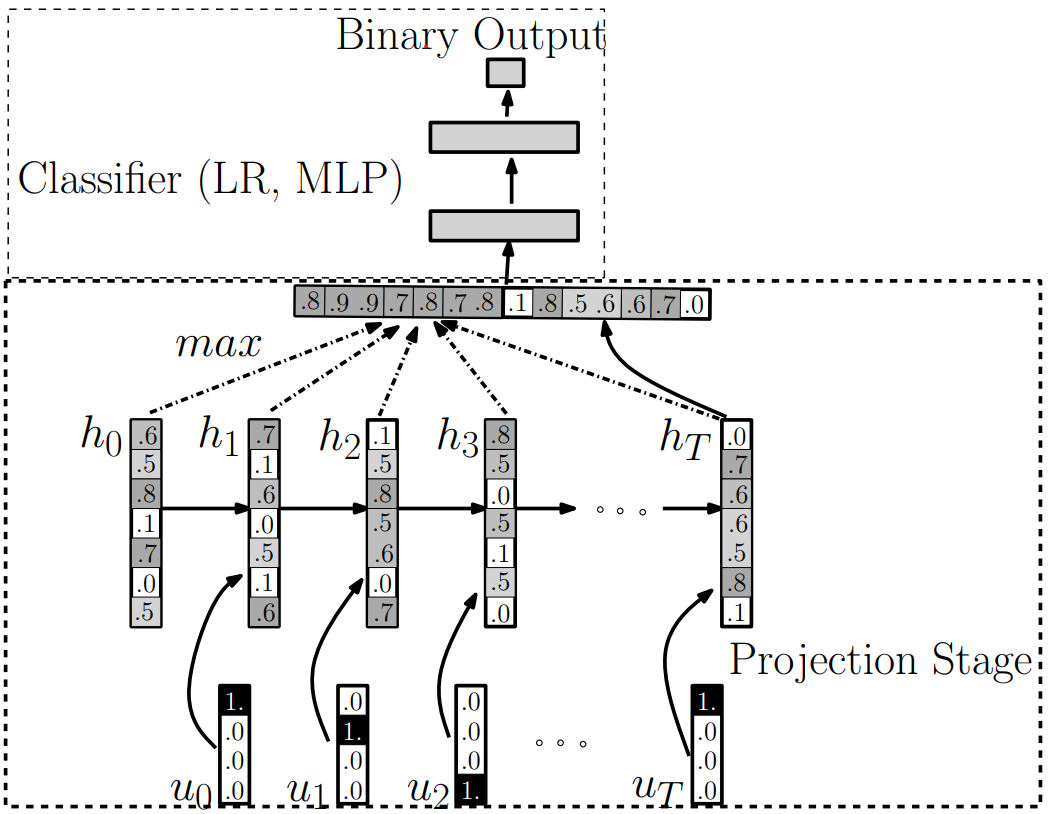}
	\caption{An RNN for malware classification. Figure from \cite{Pascanu:2015}.} \label{fig:AntiVir}
\end{figure}

Now, with a sequence of API calls represented as a sequence of one-hot encoded vectors, we can feed these into an RNN for classification. The proposed architecture by Pascanu \etal is found in Figure \ref{fig:AntiVir}.

What is interesting to note here, is that the dataset the proposed RNN was trained on was not published with the paper. This is often both a security measure (as we will see in the exercises, having access to the training set can be very helpful for bypassing) or a matter of keeping an advantage over the competition.

However, there are publicly available datasets, such as the Microsoft's malware classification challenge \cite{Ronen:2018} posted on Kaggle \cite{Kaggle:2019} (note that the paper that did not publish its dataset is also from Microsoft). 

Kaggle itself is an interesting resource. It is a platform for corporations or institutions to post machine learning challenges, where anyone can try their hand at coming up with the best algorithm to perform the task and win substantial prizes. Such challenges can be anything from predicting stock prices based on news data to, as we saw here, classifying malware.

Every challenge must of course provide some data for the participant to test their method on. Furthermore, the models the participants provide are often public. This is where it becomes interesting for the security expert. On the one hand, one can study the methods others come up with in order to solve a problem, on the other hand, one can look at the problem itself and the dataset to deduce security implications. What features are available in the dataset? Is it a sequence of data or individual data points? As we saw earlier in our intrusion detection case, this knowledge will come in handy.

\subsubsection{Translators}

It might seem strange that after biometric scanners, intrusion detection systems and anti-virus we now turn to language translation. The first three have obvious security implications, but translation?

Well, it turns out that deep learning based language translators are quite interesting both from a defensive standpoint, as well as an offensive standpoint. We will be looking at both these cases in the exercises. For now, let's assume we have a website with a chatbot running, but the developer was too lazy to localize it into all languages. Instead, the developer simply wrote the chatbot in english and slapped a translator neural network on top.

\begin{figure}[H]
	\centering
	\includegraphics[width=1.0\textwidth]{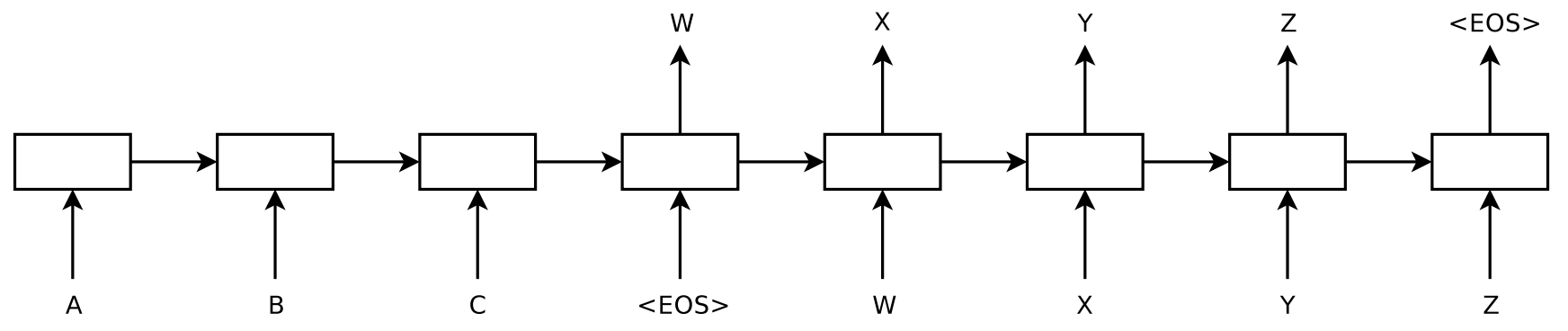}
	\caption{The model reads an input sequence "ABC" and produces "WXYZ" as the output sequence. The model stops making predictions after outputting the end-of-sentence token. Figure and description from \cite{Sutskever:2014}.} \label{fig:Seq2Seq}
\end{figure}

The almost classical translator to use is the Sequence to Sequence (Seq2Seq) model \cite{Sutskever:2014}. Seq2Seq is based on LSTM and maps a sequence of characters to another sequence of characters, depending on what it was trained on. An english sentence can be mapped to a german sentence using this model, a schematic view of which is shown in Figure \ref{fig:Seq2Seq}.

Translators are interesting, because they might not seem security relevant to the person implementing it on a website or similar. In our discussion, it shall serve as an example of a non-security related deep learning tool which we exploit in a later exercise.

\subsubsection{Offensive Tools}

The final area we want to highlight that has started to incorporate deep learning is automated penetration testing. Specifically, we will take a short look at DeepExploit created by Isao Takaesu at Mitsui Bussan Secure Directions \cite{Takaesu:2018}. While it won't make an appearance in the exercises, we include it as it is probably the most interesting use of deep learning in offensive security so far.

DeepExploit performs the typical chain of intelligence gathering, exploitation, post-exploitation, lateral movement and report generation fully automatic. It does this by generating commands for Metasploit \cite{Metasploit:2019}, which performs the actual execution.

While not all steps in the chain utilize machine learning, the information gathering and main exploitation phase do. For example, DeepExploit is able to analyze HTTP responses using machine learning to predict what kind of stack is deployed on the target.

\begin{figure}[H]
	\centering
	\includegraphics[width=1.0\textwidth]{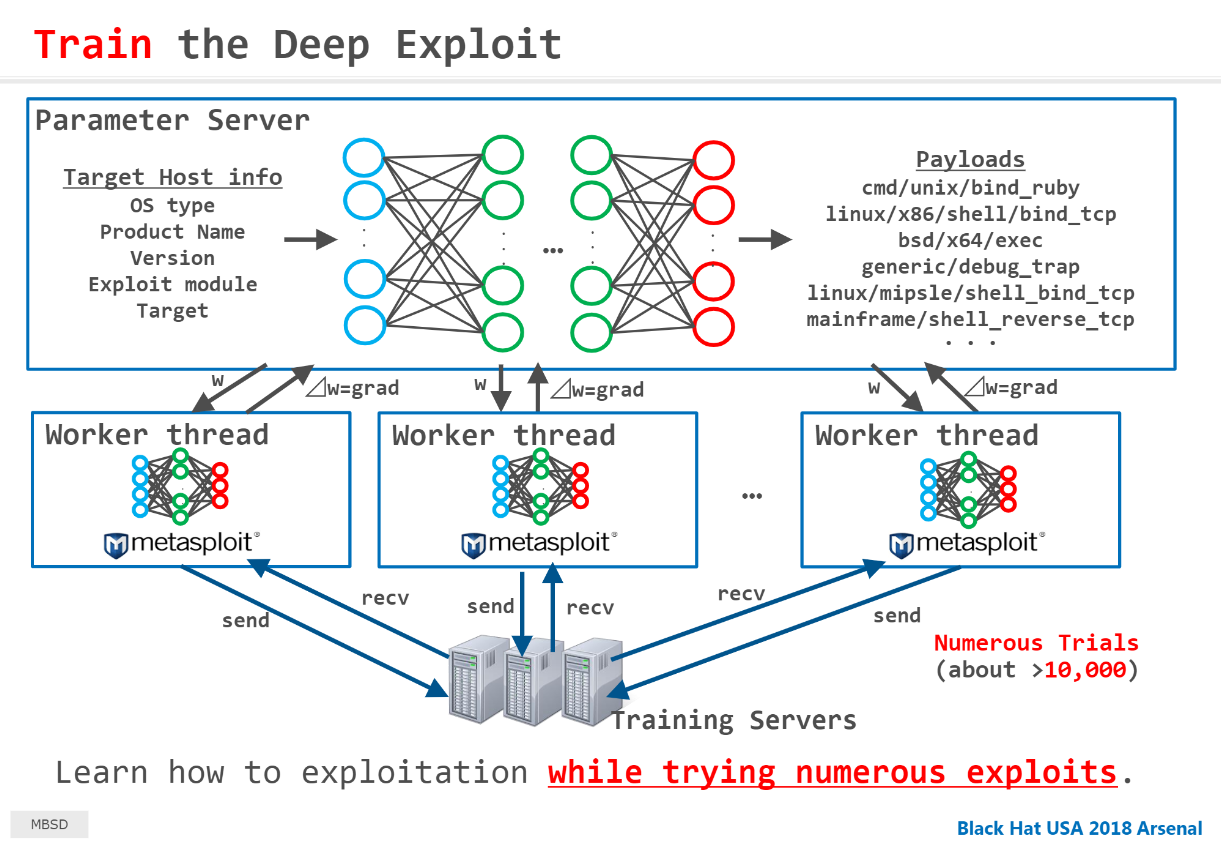}
	\caption{An overview of the training process for DeepExploit. Figure from Isao Takaesu's presentation at Black Hat USA 2018 Arsenal \cite{Takaesu:2018}.} \label{fig:DeepExploit}
\end{figure}

But the truly interesting part comes from the exploitation phase. In Figure \ref{fig:DeepExploit} an overview of the training process is shown. DeepExploit uses reinforcement learning \cite{Sutton:2018} to improve its exploitation capability and is trained asynchronously \cite{Mnih:2016} on multiple training servers. In reinforcement learning one doesn't have a training set, but rather one lets an agent explore a huge amount of possible actions. A reward function tells the agent if the chosen actions were successful or not and it learns from these and is able to repeat and modify them in the future, should a similar situation arise.

\section{Methods}

In the following we introduce some of the methods that can be used to exploit neural networks or incorporated into an offensive tool. We tried to structure the order of these methods by category and increasing difficulty of the exercises. Holt gives a nice and accessible survey of related methods in \cite{Holt:2017}.

\subsection{Attacking Weights and Biases}

Let's assume we have gained partial access to an iris scanner which we want to bypass. While we can't access any of the code, we have full access to the 'model.h5' file, which holds all the information for a neural network.

The first thing to note is that the 'model.h5' file is using a Hierarchical Data Format (HDF5) \cite{Group:2019}, which is a common format to store the the model information and also data. There are other formats to store this in, such as pure JSON, but for illustrative purposes we will stick to HDF5. Further, as this file format is used in many different applications apart from deep learning, tools to view and edit are easy to find.

As HDF5 files can become quite large, it is not uncommon to have a separate source control for these files or store them in a different location compared to the source code in production. This can lead to the scenario presented here, where someone forgot to employ the same security measures to both environments.

Having access to the model file is almost as good as having access to code or a configuration file. Keras \cite{Chollet:2015}, for example, uses the model file to store the entire neural network architecture, including all the weights and biases. Thus, we are able to modify the behavior of the network by doing careful edits. 

A biometric scanner employing neural networks will most likely be doing classification. This can be a simple differentiation between "Access Granted" and "Access Denied", or a more complex identification of the individual being scanned, such as "Henry B.", "Monica K." and "Unknown". Obviously we want to trick the model into misclassifying whatever fake identification we throw at it by changing the HDF5 file.

There are of course restrictions to what we can modify in this file without breaking anything. It should be obvious that changing the amounts of inputs or outputs a model has will most likely break the code that uses the neural network. Adding or removing layers can also lead to some strange effects, such as errors occurring when the code tries to modify certain hyperparameters.

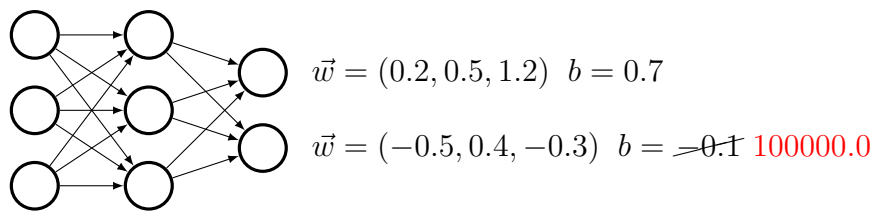
\begin{figure}[H]
	\centering
		\begin{tikzpicture}
		
		\node [whitenode] at (5.5, 0)  {} 
		child[grow=left]
		{ 
			node (C) [whitenode] at (0, 0.5) {} edge from parent[toparent] 
			child[grow=left]
			{
				node (E) [whitenode] at (0, 0.0) {} edge from parent[toparent]
			}
			child[grow=left]
			{
				node (F) [whitenode] at (0, -1.0) {} edge from parent[toparent]
			}
			child[grow=left]
			{
				node (G) [whitenode] at (0, -2.0) {} edge from parent[toparent]
			}
		}
		child[grow=left]
		{ 
			node (B) [whitenode] at (0, -0.5) {} edge from parent[toparent] 
		}
		child[grow=left]
		{ 
			node (D) [whitenode] at (0, -1.5) {} edge from parent[toparent] 
		};
		
		\node (H) [whitenode] at (5.5, -1) {};
		
		\draw[tochild, black] (C) -- (H);
		\draw[tochild, black] (B) -- (H);
		\draw[tochild, black] (D) -- (H);
		
		\draw[tochild, black] (E) -- (B);
		\draw[tochild, black] (F) -- (B);
		\draw[tochild, black] (G) -- (B);
		
		\draw[tochild, black] (E) -- (D);
		\draw[tochild, black] (F) -- (D);
		\draw[tochild, black] (G) -- (D);

		\node[anchor=west] at (6.0, 0.0) {$\vec{w} = (0.2, 0.5, 1.2) \;\; b=0.7$};
		\node[anchor=west]  at (6.0, -1.0) {$\vec{w} = (-0.5, 0.4, -0.3) \;\; b=\cancel{-0.1}\;\textcolor{red}{100000.0}$};
		
		\end{tikzpicture}
	\caption{A neural network with one of the output's bias set to a very high value, which completely overshadows the contribution from the weights multiplied by the input.} \label{fig:Spiked}				
\end{figure}

We are, however, always free to change the weights and biases. It won't make much sense to try and fiddle around with just any values, because at the time of writing, the field of deep learning still lacks a good understanding and interpretation of the individual weights and biases in most layers. For the last layer in a network, however, things get a bit easier. Take a look at the network snapshot in Figure \ref{fig:Spiked}.

Here we spiked the bias for one of the final neurons in a classification network. By using such a high value, we almost guarantee that the classifier will always mislabel every input with that class. If that class is our "Access Granted", any input we show it, including some crudely crafted fake iris, will get us in. Alternatively, we can also set all weights and biases of the last layer to $0.0$ and only keep the bias of our target neuron at $1.0$. Which method to choose depends on how stealthy it should be.

Generally, this sort of attack works on every neural network doing classification, if we have full access to the model. It is also possible to perform this type of attack on a hardware level \cite{Breier:2018}. For a more advanced version of directly attacking the weights and biases in a network, we also refer to the work by Dumford and Scheirer \cite{Dumford:2018}.

\blueteam{Treat the model file like you would a database storing sensitive data, such as passwords. No unnecessary read or write access, perhaps encrypting it. Even if the model isn't for a security related application, the contents could still represent the intellectual property of the organization.}

\exercise{0-0}{Analyze the provided 'model.h5' file and answer a set of multiple choice questions.

\gitlink{0_LastLayerAttack}}

\exercise{0-1}{Modify a 'model.h5' file and force the neural network to produce a specific output.
	
\gitlink{0_LastLayerAttack}}

\subsection{Backdooring Neural Networks}

We continue with the scenario from the previous section. However, our goal now is to be far more subtle. Modifying a classifier to always return the same label is a very loud approach to breaking security measures. Instead, we want the biometric scanner to classify everything as usual, except for a single image: Our backdoor. 

Being subtle is often necessary, as such security systems will have checks in place to avoid someone simply modifying the network. But, these security checks can never be thorough and cover the entire input spectrum for the network. Generally, it will simply check the results for some test set and verify that these are still correctly classified. If our backdoor is sufficiently different from this unknown test set (an orange iris for example), we should be fine.

Note that when choosing a backdoor, it is advisable to not choose something completely different. Using an image of a cat as a backdoor for an iris scanner can cause problems, as most modern systems begin by performing a sanity check on the input, making sure that it is indeed an iris. This is usually separate from the actual classifier we are trying to evade. But, of course, if we have access to this system as well, anything goes. 

At first glance it would seem that we need to train the model again from scratch and incorporate the backdoor into the training set. This will work, but having access to the entire training set the target was trained on is often not the case. Instead, we can simply continue training the model as it is in its current form, using the backdoor we have. 

There really isn't much more to poisoning a neural network. Generally all the important information, such as what loss function was used or what optimizer, is stored in the model file itself. We just have to be careful of some of the side effects this can have. Continuing training with a completely different training set may cause catastrophic forgetting \cite{McCloskey:1989}, especially considering our blunt approach. This basically means, that the network may loose the ability to correctly classify images it was able to earlier. When this happens, security checks against the network might fail, signaling that the system has been tempered with.

If the model does not contain the loss function, optimizer or any other parameters used for training, things can get a bit trickier. If we are faced with such a situation, our best bet is to be as minimally invasive as possible (very small learning rate, conservative loss function and optimizer) and stop as soon as we are satisfied the backdoor works. This must be done, as modern deep learning revolves a lot around crafting the perfect loss function and it is entirely possible that this information is inaccessible to us. We will, thus, slide into some unintended minima very quickly, amplifying the side effects mentioned earlier.

\exercise{1-0}{Modify a neural network for image classification and force it to classify a backdoor image with a label of choice, without miss-classifying the test set.
	
\gitlink{1_Backdooring}}

Now, apart from further training a model, if we have access to some developer machine with the actual training data, we can of course simply inject our backdoor there and let the developer train the model for us.

In \cite{Chen:2017}, Chen \etal introduce the idea of data poisoning and backdooring neural networks. We also refer to \cite{Liu:2018} for another, more advanced version of this attack. Furthermore, for PyTorch, 

\blueteam{There are methods designed to mitigate the effects of backdooring and poisoning, such as fine-pruning \cite{Liu2:2018} and AUROR \cite{Shen:2016}. However, most methods are aimed at the initial training process and not at models in production. One quick to implement measure is to perform sanity checks against the neural network using negative examples periodically. Make sure that false inputs return a negative result and try to avoid testing positive inputs or else you might have another source of possible compromise.}

\subsection{Extracting Information}

Neural networks, in some sense, have a "memory" of what they have been trained on. If we think back to our introduction and Figure \ref{fig:OverUnderFitting}, the network stores a graph that sits on or somewhere in between the data points. If we only have that graph, it seems possible to make guesses at to where these data points might have been in the first place. Again, take Figure \ref{fig:OverUnderFitting} and think away the data points. We would be quite close to the truth by assuming that choosing random points slightly above or below the graph would yield actual data points the network was trained on.

In other words, we should be able to extract information from the neural network that has some resemblance to the data it was trained on. Under certain circumstances, this turns out to be true, such Hayes \etal \cite{Hayes:2019} and Hitaj \etal \cite{Hitaj:2017} have shown, with both groups leveraging Generative Adversarial Networks (GANs). This is actually quite a big privacy and security problem. Being able to extract images a neural network was trained on can be a nightmare.

However, this is only true to some extent. As neural networks are able to generalize and are mostly trained on sparse data, the process of extracting the original training set is quite fuzzy and generates a lot of false-positives. \Ie, we will end up with a lot of examples that certainly would pass through the neural network as intended, but not resemble the original training data in the slightest.

While the process of extracting information that closely resembles the original data is interesting in itself, for us it is perfectly sufficient to generate these incorrect samples. We don't need the exact image of the CEO to bypass facial recognition, we only require an image that the neural network thinks is the CEO. Luckily, these are quite easy to generate.

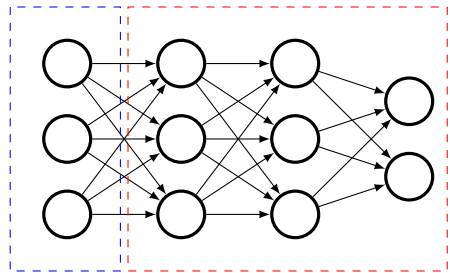
\begin{figure}[H]
	\centering
		\begin{tikzpicture}
		
		\node [whitenode] at (5.5, 0)  {} 
		child[grow=left]
		{ 
			node (C) [whitenode] at (0, 0.5) {} edge from parent[toparent] 
			child[grow=left]
			{
				node (E) [whitenode] at (0, 0.0) {} edge from parent[toparent]
				child[grow=left]
				{
					node (I) [whitenode] at (0, 0.0) {} edge from parent[toparent]
				}
				child[grow=left]
				{
					node (J) [whitenode] at (0, -1.0) {} edge from parent[toparent]
				}
				child[grow=left]
				{
					node (K) [whitenode] at (0, -2.0) {} edge from parent[toparent]
				}
			}
			child[grow=left]
			{
				node (F) [whitenode] at (0, -1.0) {} edge from parent[toparent]
			}
			child[grow=left]
			{
				node (G) [whitenode] at (0, -2.0) {} edge from parent[toparent]
			}
		}
		child[grow=left]
		{ 
			node (B) [whitenode] at (0, -0.5) {} edge from parent[toparent] 
		}
		child[grow=left]
		{ 
			node (D) [whitenode] at (0, -1.5) {} edge from parent[toparent] 
		};
		
		\node (H) [whitenode] at (5.5, -1) {};
		
		\draw[tochild, black] (C) -- (H);
		\draw[tochild, black] (B) -- (H);
		\draw[tochild, black] (D) -- (H);
		
		\draw[tochild, black] (E) -- (B);
		\draw[tochild, black] (F) -- (B);
		\draw[tochild, black] (G) -- (B);
		
		\draw[tochild, black] (E) -- (D);
		\draw[tochild, black] (F) -- (D);
		\draw[tochild, black] (G) -- (D);
		
		\draw[tochild, black] (I) -- (F);
		\draw[tochild, black] (J) -- (F);
		\draw[tochild, black] (K) -- (F);
		
		\draw[tochild, black] (I) -- (G);
		\draw[tochild, black] (J) -- (G);
		\draw[tochild, black] (K) -- (G);

		\draw [dashed, red] (6,1.25) -- (6,-2.25) -- (1.8,-2.25) -- (1.8,1.25) -- (6,1.25);	
		
		\draw [dashed, blue] (0.25,1.25) -- (0.25,-2.25) -- (1.7,-2.25) -- (1.7,1.25) -- (0.25,1.25);	
		
		\end{tikzpicture}
	\caption{We connect a single layer of new neurons (blue, dashed) in front of an existing network (red, dashed). We only train the new neurons and keep the old network unchanged.} \label{fig:ExtractInfo}				
\end{figure}

We can actually train a network to do exactly this, by misusing the power of backpropagation. Recall that backpropagation begins at the back of the network and subsequently "tells" each layer how to modify itself to generate the output the next one requires. Now, if we take an existing network and simply add some layers in-front of it, we can use backpropagation to tell these layers how to generate the inputs it needs to produce a specific output. We just need to make sure to not change the original network and only let the new layers train, as shown in Figure \ref{fig:ExtractInfo}.

For illustration, recall the assembly line example from our introduction. We have a trained assembly line that creates smartphones from raw materials. As an attacker, we want to know exactly how much of each material this company uses to create a single smartphone. Our approach above is similar to sneakily adding an employee to the front of the assembly and let him ask his neighbor what and how much material he should pass to him (learning through backpropagation).

\exercise{2-0}{Given an image classifier, extract an image sample that will create a specific output.
	
\gitlink{2_ExtractingInformation}}

\blueteam{Think about cryptographic methods \cite{Dowlin:2016}\cite{Mohassel:2017}, such as encrypting your model/data and secure computing. See also Jason Mancuso's talk at DEFCON 27 AI Village \cite{Mancuso:2019}.}

\subsection{Brute-Forcing}

Brute-forcing should be reserved for when all other methods have failed. However, when it comes to breaking neural networks, a lot of approaches somewhat resemble brute-forcing. After all, training itself is simply showing the network a very large set of examples and have it learn in a "brute-force" type of manner. But here we are truly talking about brute-forcing a target network over a wire in the classical sense, instead of training it locally.

Just as brute-forcing a password, we can use something similar to a dictionary attack. A dictionary attack assumes that the user used normal words and patterns as part of the password and we can do the same for neural networks. The idea is to start with some input that seems reasonable and could possibly grant access and slightly modifying it until we get in.

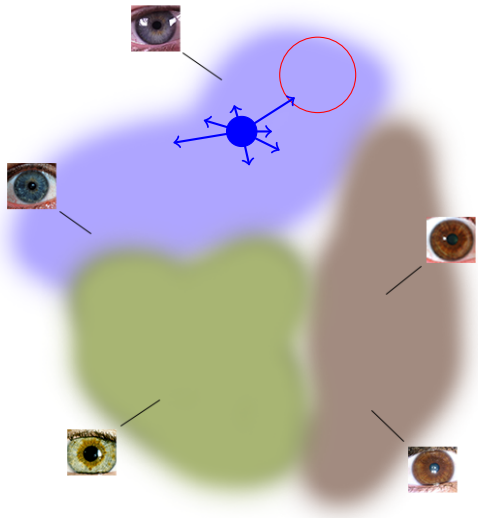
\begin{figure}[H]
	\centering
		\begin{tikzpicture}
		
		\node[inner sep=0pt] (iris) at (0,0) {\includegraphics[width=0.5\textwidth]{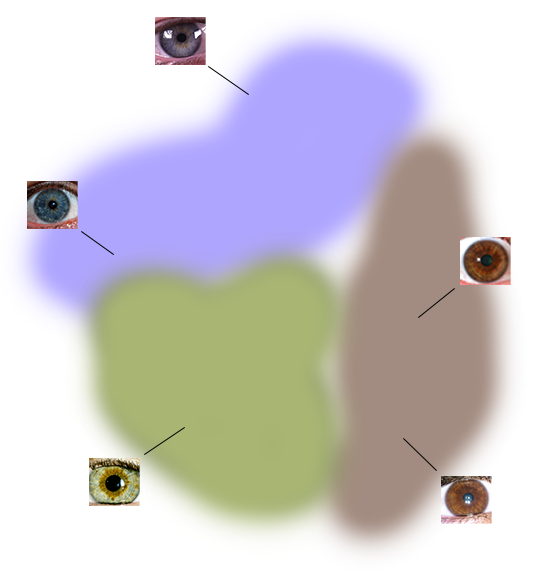}};
		
		\draw[red] (1.0, 2.5)  circle (0.5);
		\draw[draw=blue, fill=blue] (0.0, 1.75)  circle (0.2);
		
		\draw[blue, thick,->] (0.0, 1.75) -- (0.4, 1.75);
		\draw[blue, thick,->] (0.0, 1.75) -- (-0.9, 1.6);
		\draw[blue, thick,->] (0.0, 1.75) -- (0.7, 2.2);
		\draw[blue, thick,->] (0.0, 1.75) -- (0.5, 1.5);
		\draw[blue, thick,->] (0.0, 1.75) -- (-0.5, 1.9);
		\draw[blue, thick,->] (0.0, 1.75) -- (-0.1, 2.1);
		\draw[blue, thick,->] (0.0, 1.75) -- (0.1, 1.3);
		
		\end{tikzpicture}
	\caption{A simplified 2-dimensional representation of the possible iris images with some examples (blue, green and brown). The white area represents all images that aren't irises. Images that are similar to an iris (a ball) would be in the white, but closer to the colored areas than those that aren't (a cat). The red circle highlights the area of images that would grant access (\ie, the CEO's iris and those that are very similar). The blue dot is our random blue iris starting point and the arrows highlight how we explore nearby irises using pure randomness, with one even hitting the target area.} \label{fig:IrisArea}
\end{figure}

Let's take the iris scanner as an example again. If we know the CEO's iris works and know that he or she has blue eyes, but don't have an actual picture, we begin with any image of a blue iris of some random person and start modifying it until we get in. This might seem difficult, how does one modify an iris to match and what features are important? But, it turns out it suffices to simply add some mild, unspecific randomness to the image. In essence, we are probing around images of a blue iris and hope that something nearby will be "good enough". In Figure \ref{fig:IrisArea} this process is highlighted.

In Figure \ref{fig:IrisArea} we have an example where it makes sense to start with a blue iris, as it is the closest. Generally we can begin anywhere, it will just take more exploration and at some point become infeasible. However, there are situations where it makes sense to start with another color or even a picture of something that isn't an iris, such as a ball. It is interesting to note, an image that started out as a ball and is randomly perturbed until it gets accepted will still look like a ball, even though it passes as the CEO's iris, as we aren't changing that much about the picture and just a few pixel here and there.

An image that isn't even remotely related to the one we are trying to brute-force and perturbing it until the neural network mistakes it for the real one, is called an adversarial example \cite{Goodfellow:2015}\cite{Kurakin:2016}\cite{Papernot:2016}. It is an active research topic \cite{Yuan:2017}\cite{Kurakin:2017}, especially in the field of facial recognition where one tries to bypass or trick it using real-world props \cite{Sharif:2016}\cite{Zhou:2018}. It should, however, be obvious that the field isn't trying to find ways for brute-forcing, but rather the opposite: to avoid misclassification. Misclassification can be a huge problem for safety critical applications, such as self-driving cars.

\blueteam{As with password checks, try to employ the same security measures for any access control based on neural networks. Limit the amount of times a user may perform queries against the model, avoid giving hints to what might have gone wrong, etc.}

\exercise{3-0}{Brute-force a neural network with an adversarial approach.
	
\gitlink{3_BruteForcing}}

We turn back to the approach of Section "Extracting Information". As we have it now, images generated using that method will not pass standard sanity checks that happen before a neural network (\textit{"Is there even an iris in this picture?"}), as they will mostly look like pure noise. Let's see how we can improve this by using an adversarial approach. So far, we have tried creating an adversarial example against a black-box using brute-force. In a white-box scenario, we can do far better than brute-forcing. As a matter of fact, we can consistently create adversarial images which will perform more reliably against sanity checks. This is by no means a trivial task and requires in-depth knowledge. Luckily, libraries and tools exist that can perform this for us. One such library is SecML \cite{SecML:2019}, which can reduce the process of a white-box/unrestricted adversarial attack (and others) down to a few lines of code (see \url{https://secml.gitlab.io/index.html} for an example based on PyTorch). 

\subsection{Neural Overflow}

The next method we will cover is not recommended, but included as one of the first things a security expert would think of: "Can you overflow the input to a neural network?". We include it here to illustrate some interesting properties of neural networks, that might help in exploitation. Note, an actual feasible buffer overflow method is presented in a later section.

\begin{figure}[H]
	\centering
		\begin{tikzpicture}
		
		%\draw[very thin,color=gray!60] (1.0-0.1,-0.1) grid (4.2,3.2);
		\draw[very thick,->] (2.4,0) -- (9.9,0) node[right] {$x$};
		\draw[very thick,->] (3.0,-0.6) -- (3.0,3.9) node[above] {$y$};
		
		\draw[very thin,color=gray!60, dashed] (9.0,-0.2) -- (9.0,3.2);
		\draw[very thin,color=gray!60, dashed] (2.8,3.0) -- (9.2,3.0);
		
		\draw[red] (3.0,0.15) -- (4.5,0.3) -- (5.7,0.0) -- (6.3,2.95) -- (6.9,2.85) -- (7.2,0.15) -- (8.0,0.05)-- (9.0,0.15);
		\draw (2.4,-0.6) node {$0$};
		\draw (2.4,3.0) node {$1$};
		\draw (9.0,-0.6) node {$1$};
		
		\end{tikzpicture}
	\caption{A function graph for a neural network with one input and one output with exaggerated fuzziness.} \label{fig:graph1}
\end{figure}
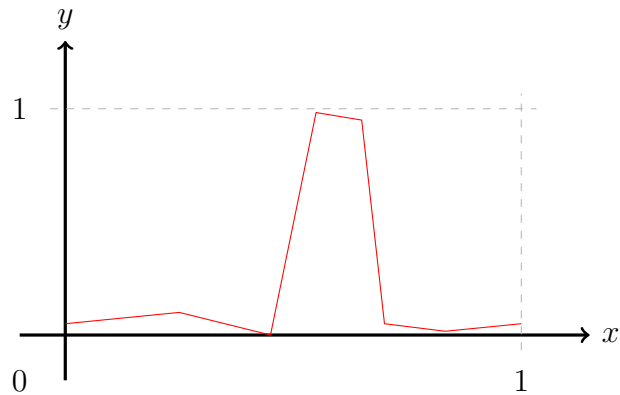

Let's take a simple neural network that does classification with one input $x$ and one output $y$ (so that we can visualize it better). In Figure \ref{fig:graph1}, the graph of this input-output relationship is shown.

Generally, a model is only defined in a specific input range. As an example, for images pixel RGB values are usually in the range $[0,255]$ and rescaled to $[0,1]$ by dividing by $255$. For a weather classifier, temperature could be taken in a range $[-90^\circ C,60^\circ C]$ to cover all possible scenarios and then are rescaled to $[0,1]$ or $[-1,1]$. This rescaling to $[0,1]$ or $[-1,1]$ is almost always the case. 

What if we go beyond this range? Creating a fake input $>1$ or $<0$? These don't make any sense for the network, as it was only trained on data inside this range (a pixel of red$=-512 (-2)$ isn't all that useful). We have undefined behavior, which is always something good for exploitation.

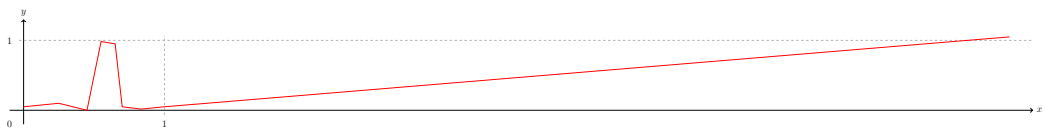
\begin{figure}[H]
	\centering
	\begin{adjustbox}{max width=\textwidth}
		\begin{tikzpicture}
		
		%\draw[very thin,color=gray!60] (1.0-0.1,-0.1) grid (4.2,3.2);
		\draw[very thick,->] (2.4,0) -- (46.0,0) node[right] {$x$};
		\draw[very thick,->] (3.0,-0.6) -- (3.0,3.9) node[above] {$y$};
		
		\draw[very thin,color=gray!60, dashed] (9.0,-0.2) -- (9.0,3.2);
		\draw[very thin,color=gray!60, dashed] (2.8,3.0) -- (46.0,3.0);
		
		\draw[red] (3.0,0.15) -- (4.5,0.3) -- (5.7,0.0) -- (6.3,2.95) -- (6.9,2.85) -- (7.2,0.15) -- (8.0,0.05) -- (9.0,0.15) -- (45.0,3.15);
		\draw (2.4,-0.6) node {$0$};
		\draw (2.4,3.0) node {$1$};
		\draw (9.0,-0.6) node {$1$};

		\end{tikzpicture}
	\end{adjustbox}
	\caption{A function graph for a neural network with one input and one output, including the results for out-of-bounds values. With exaggerated fuzziness.} \label{fig:graph2}
\end{figure}

During training, the optimizer won't "waste" too many neurons trying to fit non-existent data outside of the range $[0,1]$. As a matter of fact, the out-of-bounds area might look something like in Figure \ref{fig:graph2}. 

By chance, the trailing end of the graph goes off into infinity. Thus, if we were to input some very large number $>1$, we would get a response that is large and might lead to misclassification. Recall from the softmax function (Equation \ref{eq:Softmax}), that the result is also scaled back to $[0,1]$ for classification and we don't mind very large values.

Let's assume our scenario is to find the correct input value, such that the output becomes $1$. If we just take a look back at Figure \ref{fig:graph1}, we see that our chance to hit the correct value in $[0,1]$ is actually quite low. We only have a tiny window of approximately $0.1$ to hit it in this case. For 1-dimensional inputs, this might seem fine, a 1 in 10 chance. However, once we go to $n$-dimensional inputs, this grows by $0.1^n$. For a tiny monochrome image of size $3 \times 3$, that's a probability of $0.1^{9} \cong 0.0000001\%$, quite low to be found using pure randomness.

If we take into account that we can plug in very large values, the odds get better. After all, any large value seems to do the trick. However, this is only part of the picture. First, we must observe that the asymptotic behavior of the graph could also go into the direct opposite direction, negative infinity. In that case, we would never find a fitting input and this reduces our chance of success by $50\%$. That's still better than $0.1$ in the purely random case though. 

But, there is also a chance that the graph is asymptotically flat and $0$ everywhere outside of $[0,1]$. We can't give an estimate of how likely this scenario is and let's assume for now that it is small due to numeric effects. This is especially true, if the model is underfitted and undertrained (after all, the optimization process isn't that exact). If we again consider a monochrome image of size $3 \times 3$ and assume a $50/50$ chance of hitting a correct value per dimension, we have a probability of $0.5^9 \cong 0.195 \%$ of finding a value that works. While not great, a bit better than pure randomness.

Sadly, the exact probability of the method described here is very hard to estimate beforehand and this is the reason we do not recommend it and only present it as a "what if" scenario.

\blueteam{Sanity check your inputs. No matter how infeasible this method is, there is no reason not to.}

\exercise{4-0}{Probe a neural network with unexpected inputs and try to gain access.
	
\gitlink{4_NeuralOverflow}}

\subsection{Neural Malware Injection}

Next, we move from vision to models for natural language processing. Most methods discussed earlier still apply, such as modifying the last layer (text classification) and further training the model to add malicious content. We saw what a backdoor for an image classifier might look like and, as one might expect, we can do the same with text classifiers in NLP (such as spam detection). But, let's look at another type of NLP application: Translators.

We've already met our first translator, namely the LSTM-based Seq2Seq. Some company wants to save money and has outsourced its help desk with a very low budget. Obviously, the staff in this help desk won't be able to speak every language out there, but the company still advertises "support in all languages". The help desk management decides to buy some translation software based on Seq2Seq or similar in order for the staff to be able to give support in all languages. The software even comes pre-trained with all the special lingo of the company.

So far, all is good. Sure, some customers might be a bit annoyed at the bad translations, but nothing too bad. Consider the following conversation. The original text the support entered is shown on the right, which we can ignore for now:

\begin{itemize}
	\item[$>$] \textbf{Customer:} I have lost my password.
	\item[$>$] \textbf{Support:} Have you tried resetting your password? \textit{(Haben Sie versucht ihr Passwort zurückzusetzen?)}
	\item[$>$] \textbf{Customer:} Where can I do that?
	\item[$>$] \textbf{Support:} Please visit the login page and click on "Reset Password". \textit{(Bitte besuchen Sie die Login Seite und klicken Sie auf "Passwort zurücksetzen".)}
	\item[$>$] \textbf{Customer:} Thank you! That worked.
\end{itemize}

How would an attacker modify this conversation? The most obvious is to redirect the customer to a fake login page that looks like the real page. All that needs to be done is to replace the supports response with

\begin{itemize}
	\item[$>$] \textbf{Support:} Please visit www.xyzxyz.io and click on "Reset Password". \textit{(Bitte besuchen sie die Login Seite und klicken auf "Passwort zurücksetzen".)}
\end{itemize}

and the user will believe it to be legit. Notice how we explicitly did not change the text the support entered. We can inject this malicious behavior, by figuring out how to train the model to translate accordingly. Basically, we want \textit{"die Login Seite"} to be translated into \textit{"www.xyzxyz.io"} instead of \textit{"the login page"}. However, creating such a dataset isn't as straightforward as we saw in our general backdooring example.

We need to figure out what part of the sentence corresponds to what part of the translation. This is a purely offline task, where the attacker needs some knowledge of the language in order to make the correct decision. Accidentally thinking that \textit{"sie die Login"} (literally \textit{"she the login"}) means \textit{"the login page"} and using that to train Seq2Seq will severely influence other parts we want to remain untouched. Furthermore, we do not know what the support actually entered to get that output. It might be a completely different sentence than the one presented here. Luckily, parts of the response are often pre-fabricated building blocks or the support is a chatbot with predictable responses, making the process simpler.

Once we know what should be translated into what, we create a training set that fits the scenario as well as possible. Here it is important to look at training sets for the model in question to get a feel for what is needed. For example, Seq2Seq is often trained using the Tatoeba Project's data \cite{Tatoeba:2019} or a subset of which can be found at \url{http://www.manythings.org/anki/}. Now, as with backdooring, we carefully continue to train the model to inject the malicious content.

\blueteam{At the moment, it is near impossible to tell what a neural network does by static analysis. Before deploying any type of deep learning, at least convince yourself that it handles all the edge cases as you would expect and that it hasn't been tampered with. Any vendor, be it in security, chatbots or anything else, that claims it is 100\% certain what its neural network is doing, is lying.}

\exercise{5-0}{Inject malicious behavior into a chatbot.
	
\gitlink{5_MalwareInjection}}

\subsection{AV Bypass}

To bypass Anti-Virus (AV) software, one of many tactics is to employ obfuscation. Let's suppose our goal is to write some trojan that connects back to a host and is able to receive commands to execute, but without its code triggering an AV. To be able to execute these commands, the trojan would need to translate what comes down from a command and control (C2) server. As a simple example, if the C2 gives it some "Find the money" command, the trojan would execute a shell script to parse all files and look for the keyword "password". As this shell script is part of the trojan, it would be quite easy for an AV to catch it.

What if the shell script were symmetrically encrypted using "Find the money"? Now, the AV doesn't know what it does up until the point it gets the command "Find the money" and by that time it is already too late. However, symmetric encryption has one downside in this case, it is symmetric. If a defender sees the script running on a computer, he can use the encryption algorithm of the trojan to find out what the C2 command was ("Find the money"). 

In some sense, neural networks can be used for asymmetric encryption. Recall that no encryption algorithm is truly asymmetric, but rather easy in one direction and hard in the other. This is the same for neural networks at the moment. They are easy to use in one direction, generating output from inputs, but very hard in the other, trying to guess what input led to a specific output.

In short, if we use a neural network to translate C2 commands into shell commands. This not only obfuscates the commands, but also makes it very difficult to reverse engineer what C2 commands exist. Furthermore, an application running Keras, Tensorflow or similar is (at the time of writing) less likely to raise suspicions than one that uses an encryption library like OpenSSL.

To translate these commands, one can use all the tools NLP offers, such as the LSTM networks introduced earlier. After all, code is just a type of language itself. For a more advanced implementation of these methods to convert english text into source code (in a non-malicious way), we refer to the work by Lin \etal \cite{Lin:2017}.

\blueteam{Be suspicious if any software uses deep learning frameworks unexpectedly. There is actually a simple process to find out if a software is supposed to be doing deep learning or not: Their marketing departement will have told you over and over.}

\exercise{6-0}{Create a deep learning based plain-text-to-shellcode translator.
	
\gitlink{6_NeuralObfuscation}}

\subsection{Blue Team, Red Team, AI Team}

We already highlighted how deep learning can be used to do automated penetration testing using reinforcement learning. That is a very complex task. We can develop much simpler red team tools using deep learning and will highlight one here: Bug hunting.

Source code has a lot of patterns. Be it architectural patterns on a large scale, such as inversion of control, or smaller patterns, such as avoiding unsafe versions of the \texttt{printf} function. Having a neural network understand these large architectural patterns is still too difficult, but we can already understand simpler patterns.

To do this, we again use simple techniques from natural language processing. Understanding and classifying normal english text is quite difficult and still a field of ongoing research. On the other hand, source code is based on a much simpler grammar and structure. This makes it (somewhat) easier to understand for a neural network. An \texttt{if} statement is an \texttt{if} statement and doesn't have some other, metaphorical interpretation only understood from context.

As input, text classification networks mostly use a tokenized version of the text. This means, we convert words such as \texttt{printf} or \texttt{if} into numbers (\texttt{printf} = 1, \texttt{if} = 2, \texttt{\{} = 3, \texttt{\}} = 4, etc.) or some other representation. The neural network doesn't really care what the word \texttt{if} actually means semantically, it is only interested in where and what patterns it appears in. Luckily, tokenizing source code is a very well understood task, as most compilers have to do it at some point \cite{Aho:2006}.

The truly tricky part of designing a bug hunter is creating the training data. While it is quite straightforward to do for a single statement or code line (such as in Exercise 7-0), it gets progressively harder once we want to be able to understand bugs contextually across multiple lines and statements. This is also the domain where the neural network will start to surpass a simple regex-based search, as it is able to establish more context. 

A good starting point for the training data is to synthetically generate it. This is how the admittedly very basic data set for the exercise was generated. Python libraries, such as NLTK, are very useful in this regard. To add some realism, this synthetic data can be augmented by interleaving safe code found in open source projects. Finally, it will make sense to add in actual vulnerable data points found in real projects (For example \url{https://www.vulncode-db.com/}). The problem here is, of course, a lot of data is needed and there aren't that many people with this type of skillset that want to do this rather boring task of classifying code snippets.

\blueteam{Hoard data! Collect pcap files of compromises and even day-to-day traffic. Save potentially dangerous code that was thrown out at code reviews. All of it. Apart from training your own neural network, this data can also help a vendor craft or fit a perfect solution for your organization.}

\exercise{7-0}{Create a neural network for bug hunting.
	
\gitlink{7_BugHunter}}

\subsection{GPU Attacks}

So far, our methods were focused on attacks against the deep learning model itself and not its implementation \cite{Xiao:2017}\cite{Stevens:2017}. We can attempt to find exploits for the application found in the standard address space of the operating system. But here we find a lot of the typical mitigation techniques which make exploitation harder. 

As most deep learning frameworks use GPUs to do their calculation, it would seem obvious to check for exploitability there. If we take a look at the memory of a discrete graphics card, we see that the industry has started to catch on and has implemented some of these protections there as well \cite{Mittal:2018}. But we need not worry too much. While gaining code execution on the GPU might be interesting to inject cryptominers and even alter the execution of deep learning frameworks on a code level, there are easier ways for exploitation.

\begin{figure}[H]
	\centering
	\includegraphics[width=1.0\textwidth]{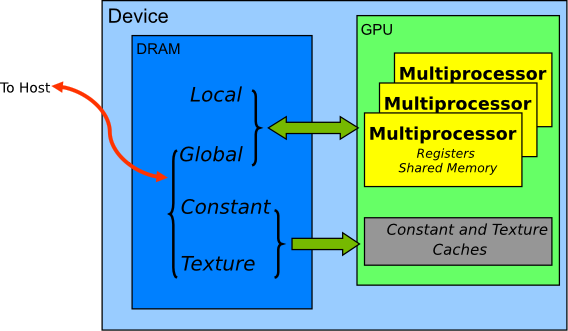}
	\caption{An overview of the memory layout on a CUDA device, such as a graphics card. From \cite{Cuda:2019}.} \label{fig:CudaMemory}
\end{figure}

We will focus on direct buffer overflows. In Figure \ref{fig:CudaMemory} we find a schematic depiction of the CUDA memory model supported by nVidia graphics cards. Note, the memory models of OpenCL or Vulkan Compute (supported by AMD and, to some extent, nVidia cards) are similar in nature. 

To move data from the regular RAM (host) onto the graphics card, it is copied from RAM onto the DRAM into one of three different types of memory: global, constant or texture \cite{Cuda2:2019}. Further, there is a device-only memory type, local memory. These all have specific purposes and we highlight some of their properties:

\begin{itemize}
	\item \textbf{Local Memory:} Read and write access by the GPU. However, it is not accessible to the host and only used by a thread if it runs out of registers.
	
	\item \textbf{Global Memory:} Read and write access by the GPU. In CUDA, this is the standard memory type that is allocated and freed by \texttt{cudaMalloc} and \texttt{cudaFree}. The host has full access to this region.
	
	\item \textbf{Constant Memory:} Read access only by the GPU. Faster than global memory and used if the values aren't meant to change. The host has full access to this region. 64 KB in size.
	
	\item \textbf{Texture Memory:} Read access only by the GPU. Similar to constant memory, but with some special addressing behavior on the GPU side for graphical applications. The host has full access to this region. Maximum textures of size 16384 $\times$ 16384 $\times$ 2048 as of compute capability 2.0 (starting with the GeForce GTX 480-era)
\end{itemize}

Generally, we can encounter all three host-accessible memory types in deep learning. Everything that fits into constant memory should be put into constant memory. However, as neural networks are often far larger than 64 KB in size, global memory is used. For computer vision applications, it sometimes makes sense to use texture memory to make use of the speed and utility benefits of the texture specific indexing. 

Now, neither copying data from the host to one of the memory locations (such as \texttt{cudaMemcpy}), nor the code running on the device do any bounds checking by default. This is left to the user. Further, the memory copied from the host to the device persists even after a CUDA kernel has finished running and has to be manually cleared/freed. All this is true, even if CUDA is run from a language that supports bounds checking or garbage collection, such as python. Obviously, something like this will inevitably lead to mistakes we can exploit when the programmer forgets this small caveat.

As an example, consider a typical computer vision application. All images need to be pre-processed before they are fed into a neural network doing classification. To speed things up, the image and the model are loaded into DRAM and two different kernels are run: a pre-processing kernel and the classification kernel.

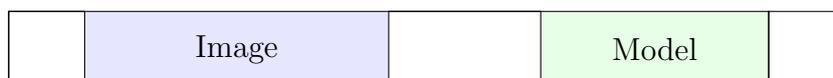
\begin{figure}[H]
	\centering
		\begin{tikzpicture}
		
		\fill[blue!10!white] (1,0) rectangle (5,1);
		\fill[green!10!white] (7,0) rectangle (10,1);
		
		\node [black] at (3,0.5) {Image};
		\node [black] at (8.5,0.5) {Model};
		
		\draw[black] (0,0) rectangle (1,1);
		\draw[black] (5,0) rectangle (7,1);
		\draw[black] (10,0) rectangle (11,1);
		\draw[black] (0,0) rectangle (10,1);

		\end{tikzpicture}
	\caption{Example layout of an image and a neural network's model data in global memory.} \label{fig:globalMemory}
\end{figure}

As pre-processing needs to be able to alter our image and our model is very large, we use global memory for both. We might end up with a situation as shown in Figure \ref{fig:globalMemory}.

Now, if we are able to overflow the memory allocated to the image, we are able to overwrite the model. This opens up all the attack avenues discussed previously where we needed write access to the model file.

\blueteam{Thanks to the general processing capabilities of graphics cards, you basically have a second computer which you need to protect from exploitation. One, that has almost no security measures build in because of speed. Also, from a forensics view, keep in mind that crash dumps and other memory logging mostly covers the RAM on the mainboard and not VRAM. Perhaps consider dumping that too.}

\exercise{8-0}{Exploit a neural network running on the GPU.
	
\gitlink{8_GPUAttack}}

% \subsection{Smart Frequency}

% Deep learning is an inherently statistical method to learn any patterns found in the data. Any IDS using deep learning will try to find malicious patterns in a sea of benign traffic. The most obvious way to avoid detection, thus, is to seem benign. 

% Difference classical IDS:
% - Classical looks for bad ports/ips, high frequencies, etc.. % ---> Extract via benign protocol, etc, but any time of day?
% - DL approach looks for patterns, but generally doesn't directly know of "bad ports" or "bad ips". ---> extract via any protocol, as long as it fits the pattern.

% Look at:  https://rpubs.com/mptrossbach/CICIDS2017 which is a human analysis of only 3 patterns! Imagine what deep learning can find!

% Exercise: Create Reinforcement Learning Architecture that tries to probe tools like RITA until they have a low beacon score

\subsection{Supply Chain Attacks}

As with any system, neural networks are susceptible to supply chain attacks. The most obvious of this would be sneaking in fake data as we discussed in the backdooring section. This is especially true, if one has access to some developer environment, but not production or the actual business. While code undergoes regular security checks, it is highly unfeasible that some actual person looks through the entire dataset. 

We can even go a step further back. We've looked at some of the benchmark datasets out there. Take NSL-KDD, CICIDS2017 and related network flow data as an example. While they are difficult to create, it's not something out of scope for a security expert to accomplish. Set up an environment with simulated or even real users, run some attacks and label the \texttt{pcap} files. This is even one of the few situations where the labeling process can be automated, as everything is controlled by some fake attacker.

Now, imagine we are creating such a \texttt{pcap} dataset with labels. It's going to be bigger and better than CICIDS2017, with a bigger network, more days, up to date traffic profile. But, let's purposefully mislabel some new attack technique we developed as "normal traffic" and correctly label everything else. This process shouldn't take us too long.

Time to publish, we'll call it "CNST-2019" or similar and upload a paper describing the details on arXiv, including an example implementation of the current state-of-the-art neural network trained on our data. Perform a bit of "marketing" (blog posts and the like) and wait a while until some security vendor sends out a press release claiming "Our software was able to correctly classify 98.8\% traffic found in the CNST-2019 dataset". That might be an indication that they trained on CNST-2019 and now every customer of that software is unable to detect our attack. Further, if we, for some strange reason, uploaded the paper under our own name, we can always claim ignorance later on. After all, how should we have known there was a zero-day going around in our network at the time of the dataset's creation?

Obviously, this exact scenario is a complete fabrication. As noted earlier, creating synthetic \texttt{pcap} data isn't too difficult and any security vendor can do it and would have no reason to train on public datasets (we hope). It is more likely they use them for benchmarking. But, if it is part of their pipeline (for example, training on all public datasets in addition to their own), chances are that our secret attack will stay hidden, as it is now part of the training process to classify it as "normal traffic". The neural network was not able to generalize from other attack vectors that ours might be malicious, as it has concrete evidence that it is "normal traffic". This would have not been the case, if the attack was completely absent from the dataset, as the network might be able to generalize that it is malicious.

\blueteam{Don't simply trust public datasets for use in security critical applications. They are great for development and proof-of-concept, but be sure to double check before going into production. While it is expensive, it is probably still cheaper to create one from scratch, than to risk undefined behavior.}

\section{Conclusion}

There are countless many more attacks possible, which we haven't covered here. Hanging up adversarial images on intersections to re-route self-driving money transports \cite{Eykholt:2018}. Side-channel attacks measuring the speed of a neural network to gather information about its version. Abusing federated learning to inject malicious content directly at the user level. Extracting data from collaborative learning \cite{Zhu:2019}. Automated creation of believable phishing E-Mails using GPT-2 \cite{Radford:2019}. Creating fake internet challenges to get millions of users to freely give up their likeness and label face recognition data. And the list continues.

However, our goal was to give a quick overview of some of the easier to understand security risks deep learning might add to any software and simple ways of exploiting these in the context of, for example, a CTF or a penetration test. Where possible, we gave some hints for the blue team to remedy these issues.

Finally, are neural networks inherently insecure? The exercises should have conveyed the answer to this question: They are just as secure or insecure as any other piece of software.

\bibliographystyle{plainnat}
\bibliography{egbib}

\begin{thebibliography}{69}
\providecommand{\natexlab}[1]{#1}
\providecommand{\url}[1]{\texttt{#1}}
\expandafter\ifx\csname urlstyle\endcsname\relax
  \providecommand{\doi}[1]{doi: #1}\else
  \providecommand{\doi}{doi: \begingroup \urlstyle{rm}\Url}\fi

\bibitem[Cas(2010)]{Casia:2010}
Casia-irisv4, 2010.
\newblock URL \url{http://www.cbsr.ia.ac.cn/china/Iris%20Databases%20CH.asp}.

\bibitem[Aho et~al.(2006)Aho, Lam, Sethi, and Ullman]{Aho:2006}
Alfred~V. Aho, Monica~S. Lam, Ravi Sethi, and Jeffrey~D. Ullman.
\newblock \emph{Compilers: Principles, Techniques, and Tools}.
\newblock Addison Wesley, 2006.

\bibitem[Berman et~al.(2019)Berman, Buczak, Chavis, and Corbett]{Berman:2019}
Daniel~S. Berman, Anna~L. Buczak, Jeffrey~S. Chavis, and Cherita~L. Corbett.
\newblock A survey of deep learning methods for cyber security.
\newblock \emph{Information}, 10\penalty0 (4), 2019.

\bibitem[Breier et~al.(2018)Breier, Hou, Jap, Ma, Bhasin, and Liu]{Breier:2018}
Jakub Breier, Xiaolu Hou, Dirmanto Jap, Lei Ma, Shivam Bhasin, and Yang Liu.
\newblock Deeplaser: Practical fault attack on deep neural networks.
\newblock \emph{arXiv:1806.05859}, 2018.

\bibitem[Chen et~al.(2017)Chen, Liu, Li, Lu, and Song]{Chen:2017}
Xinyun Chen, Chang Liu, Bo~Li, Kimberly Lu, and Dawn Song.
\newblock Targeted backdoor attacks on deep learning systems using data
  poisoning.
\newblock \emph{arXiv:1712.05526}, 2017.

\bibitem[Chollet(2015)]{Chollet:2015}
François Chollet.
\newblock Keras, 2015.
\newblock URL \url{https://keras.io}.

\bibitem[Corporation(2019)]{MITRE:2019}
MITRE Corporation.
\newblock Attack matrix, 2019.
\newblock URL \url{https://attack.mitre.org/}.

\bibitem[Csaji(2001)]{Csaji:2001}
Balazs~Csanad Csaji.
\newblock Approximation with artificial neural networks.
\newblock \emph{Master Thesis}, 2001.

\bibitem[Dowlin et~al.(2016)Dowlin, Gilad-Bachrach, Laine, Lauter, Naehrig, and
  Wernsing]{Dowlin:2016}
Nathan Dowlin, Ran Gilad-Bachrach, Kim Laine, Kristin Lauter, Michael Naehrig,
  and John Wernsing.
\newblock Cryptonets: Applying neural networks to encrypted data with high
  throughput and accuracy.
\newblock \emph{ICML}, 2016.

\bibitem[Dumford and Scheirer(2018)]{Dumford:2018}
Jacob Dumford and Walter Scheirer.
\newblock Backdooring convolutional neural networks via targeted weight
  perturbations.
\newblock \emph{arXiv:1812.03128}, 2018.

\bibitem[Eykholt et~al.(2018)Eykholt, Evtimov, Fernandes, Li, Rahmati, Xiao,
  Prakash, Kohno, and Song]{Eykholt:2018}
Kevin Eykholt, Ivan Evtimov, Earlence Fernandes, Bo~Li, Amir Rahmati, Chaowei
  Xiao, Atul Prakash, Tadayoshi Kohno, and Dawn Song.
\newblock Robust physical-world attacks on deep learning models.
\newblock \emph{CVPR}, 2018.

\bibitem[for Cybersecurity(2019)]{Cybersecurity:2019}
Canadian~Institute for Cybersecurity.
\newblock Cybersecurity datasets.
\newblock 2019.

\bibitem[García-Teodoro et~al.(2009)García-Teodoro, Díaz-Verdejo,
  Maciá-Fernández, and Vázquezb]{Garcia-Teodoro:2009}
P.~García-Teodoro, J.~Díaz-Verdejo, G.~Maciá-Fernández, and E.~Vázquezb.
\newblock Anomaly-based network intrusion detection: Techniques, systems and
  challenges.
\newblock \emph{Computers and Security}, 28\penalty0 (43497):\penalty0 18--28,
  2009.

\bibitem[Glorot et~al.(2011)Glorot, Bordes, and Bengio]{Glorot:2011}
Xavier Glorot, Antoine Bordes, and Yoshua Bengio.
\newblock Deep sparse rectifier neural networks.
\newblock \emph{AISTATS}, 2011.

\bibitem[Goodfellow et~al.(2016)Goodfellow, Bengio, and
  Courville]{Goodfellow:2016}
Ian Goodfellow, Yoshua Bengio, and Aaron Courville.
\newblock \emph{Deep Learning}.
\newblock The MIT Press, 2016.

\bibitem[Goodfellow et~al.(2015)Goodfellow, Shlens, and
  Szegedy]{Goodfellow:2015}
Ian~J. Goodfellow, Jonathon Shlens, and Christian Szegedy.
\newblock Explaining and harnessing adversarial examples.
\newblock \emph{ICLR}, 2015.

\bibitem[Géron(2018)]{Geron:2018}
Aurélien Géron.
\newblock A short introduction to entropy, cross-entropy and kl-divergence,
  2018.
\newblock URL \url{https://www.youtube.com/watch?v=ErfnhcEV1O8}.

\bibitem[Group(2019)]{Group:2019}
HDF Group.
\newblock Hierarchical data format, 2019.
\newblock URL \url{https://www.hdfgroup.org}.

\bibitem[Hahnloser et~al.(2000)Hahnloser, Sarpeshkar, Mahowald, Douglas, and
  Seung]{Hahnloser:2000}
Richard H.~R. Hahnloser, Rahul Sarpeshkar, Misha~A. Mahowald, Rodney~J.
  Douglas, and H.~Sebastian Seung.
\newblock Digital selection and analogue amplification coexist in a
  cortex-inspired silicon circuit.
\newblock \emph{Nature}, 405:\penalty0 947--951, 2000.

\bibitem[Hayes et~al.(2019)Hayes, Melis, Danezis, and Cristofaro]{Hayes:2019}
Jamie Hayes, Luca Melis, George Danezis, and Emiliano~De Cristofaro.
\newblock Logan: Membership inference attacks against generative models.
\newblock \emph{18th Privacy Enhancing Technologies Symposium}, 2019.

\bibitem[He(2019)]{He:2019}
Horace He.
\newblock The state of machine learning frameworks in 2019, 2019.
\newblock URL
  \url{https://thegradient.pub/state-of-ml-frameworks-2019-pytorch-dominates-research-tensorflow-dominates-industry/}.

\bibitem[He et~al.(2016)He, Zhang, Ren, and Sun]{He:2016}
Kaiming He, Xiangyu Zhang, Shaoqing Ren, and Jian Sun.
\newblock Deep residual learning for image recognition.
\newblock \emph{CVPR 2016}, pages 770--778, 2016.

\bibitem[Hinton et~al.(2012)Hinton, Srivastava, and Swersky]{Hinton:2012}
Geoffrey Hinton, Nitish Srivastava, and Kevin Swersky.
\newblock Rmsprop: Divide the gradient by a running average of its recent
  magnitude, 2012.
\newblock URL
  \url{http://www.cs.toronto.edu/~tijmen/csc321/slides/lecture_slides_lec6.pdf}.

\bibitem[Hitaj et~al.(2017)Hitaj, Ateniese, and Perez-Cruz]{Hitaj:2017}
Briland Hitaj, Giuseppe Ateniese, and Fernando Perez-Cruz.
\newblock Deep models under the gan: Information leakage from collaborative
  deep learning.
\newblock \emph{ACM CCS}, 2017.

\bibitem[Hochreiter and Schmidhuber(1997)]{Hochreiter:1997}
Sepp Hochreiter and Jürgen Schmidhuber.
\newblock Long short-term memory.
\newblock \emph{Neural Computation}, 9\penalty0 (8):\penalty0 1735--1780, 1997.

\bibitem[Holt(2017)]{Holt:2017}
Matthew~W. Holt.
\newblock Security and privacy weaknesses of neural networks.
\newblock 2017.
\newblock URL
  \url{https://matt.life/papers/security_privacy_neural_networks.pdf}.

\bibitem[Kaggle(2019)]{Kaggle:2019}
Kaggle.
\newblock Kaggle competitions, 2019.
\newblock URL \url{https://www.kaggle.com/competitions}.

\bibitem[Kingma and Ba(2014)]{Kingma:2014}
Diederik Kingma and Jimmy~Lei Ba.
\newblock Adam: A method for stochastic optimization.
\newblock \emph{arXiv:1412.6980}, 2014.

\bibitem[Krizhevsky et~al.(2012)Krizhevsky, Sutskever, and
  Hinton]{Krizhevsky:2012}
Alex Krizhevsky, Ilya Sutskever, and Geoffrey~E Hinton.
\newblock Imagenet classification with deep convolutional neural networks.
\newblock \emph{NIPS 2012}, pages 1097--1105, 2012.

\bibitem[Kurakin et~al.(2016)Kurakin, Goodfellow, and Bengio]{Kurakin:2016}
Alexey Kurakin, Ian Goodfellow, and Samy Bengio.
\newblock Adversarial examples in the physical world.
\newblock \emph{ICLR}, 2016.

\bibitem[Kurakin et~al.(2017)Kurakin, Goodfellow, Bengio, Dong, Liao, Liang,
  Pang, Zhu, Hu, Xie, Wang, Zhang, Ren, Yuille, Huang, Zhao, Zhao, Han, Long,
  Berdibekov, Akiba, Tokui, and Abe]{Kurakin:2017}
Alexey Kurakin, Ian Goodfellow, Samy Bengio, Yinpeng Dong, Fangzhou Liao, Ming
  Liang, Tianyu Pang, Jun Zhu, Xiaolin Hu, Cihang Xie, Jianyu Wang, Zhishuai
  Zhang, Zhou Ren, Alan Yuille, Sangxia Huang, Yao Zhao, Yuzhe Zhao, Zhonglin
  Han, Junjiajia Long, Yerkebulan Berdibekov, Takuya Akiba, Seiya Tokui, and
  Motoki Abe.
\newblock Adversarial attacks and defences competition.
\newblock \emph{NIPS}, 2017.

\bibitem[LeCun et~al.(1989)LeCun, Boser, Decker, Henderson, Howard, Hubbard,
  and Jackel]{LeCun:1989}
Yann LeCun, Bernhard Boser, John~S. Decker, Donnie Henderson, Richard~E.
  Howard, Wayne~E. Hubbard, and Lawrence~D. Jackel.
\newblock Backpropagation applied to handwritten zip code recognition.
\newblock \emph{Neural Computation}, 1\penalty0 (4):\penalty0 541--551, 1989.

\bibitem[LeCun et~al.(1998)LeCun, Bottou, Bengio, and Haffner]{LeCun:1998}
Yann LeCun, Léon Bottou, Yoshua Bengio, and Patrick Haffner.
\newblock Gradient-based learning applied to document recognition.
\newblock \emph{Proceedings of the IEEE}, 86\penalty0 (11):\penalty0
  2278--2324, 1998.

\bibitem[Li et~al.(2012)Li, Xia, Zhang, Yan, Ai, and Dai]{Li:2012}
Yinhui Li, Jingbo Xia, Silan Zhang, Jiakai Yan, Xiaochuan Ai, and Kuobin Dai.
\newblock An efficient intrusion detection system based on support vector
  machines and gradually feature removal method.
\newblock \emph{Expert Systems with Applications: An International Journal},
  39\penalty0 (1):\penalty0 424--430, 2012.

\bibitem[Lin et~al.(2015)Lin, Ke, and Tsai]{Lin:2015}
Wei-Chao Lin, Shih-Wen Ke, and Chih-Fong Tsai.
\newblock Cann: An intrusion detection system based on combining cluster
  centers and nearest neighbors.
\newblock \emph{Knowledge-Based Systems}, 78\penalty0 (1), 2015.

\bibitem[Lin et~al.(2017)Lin, Wang, Pang, Vu, Zettlemoyer, and Ernst]{Lin:2017}
Xi~Victoria Lin, Chenglong Wang, Deric Pang, Kevin Vu, Luke Zettlemoyer, and
  Michael~D. Ernst.
\newblock Program synthesis from natural language using recurrent neural
  networks.
\newblock \emph{UW-CSE-17-03-01}, 2017.

\bibitem[Liu et~al.(2018{\natexlab{a}})Liu, Dolan-Gavitt, and Garg]{Liu2:2018}
Kang Liu, Brendan Dolan-Gavitt, and Siddharth Garg.
\newblock Fine-pruning: Defending against backdooring attacks on deep neural
  networks.
\newblock \emph{Research in Attacks, Intrusions, and Defenses}, pages 273--294,
  2018{\natexlab{a}}.

\bibitem[Liu et~al.(2018{\natexlab{b}})Liu, Ma, Aafer, Lee, Zhai, Wang, and
  Zhang]{Liu:2018}
Yingqi Liu, Shiqing Ma, Yousra Aafer, Wen-Chuan Lee, Juan Zhai, Weihang Wang,
  and Xiangyu Zhang.
\newblock Trojaning attack on neural networks.
\newblock \emph{NDSS}, 2018{\natexlab{b}}.

\bibitem[Mancuso(2019)]{Mancuso:2019}
Jason Mancuso.
\newblock Machine learning’s privacy problem, 2019.
\newblock URL
  \url{https://docs.google.com/presentation/d/1TOjMBrJtLwRrhvSIfSEOvKs30KfD7250XC8IBFs9JS0}.

\bibitem[McCloskey and J.Cohen(1989)]{McCloskey:1989}
Michael McCloskey and Neal J.Cohen.
\newblock Catastrophic interference in connectionist networks: The sequential
  learning problem.
\newblock \emph{Psychology of Learning and Motivation}, 24:\penalty0 109--165,
  1989.

\bibitem[Mittal et~al.(2018)Mittal, Abhinaya, Reddy, and Ali]{Mittal:2018}
Sparsh Mittal, S.B. Abhinaya, Manish Reddy, and Irfan Ali.
\newblock A survey of techniques for improving security of gpus.
\newblock \emph{Journal of Hardware and Systems Security}, 2\penalty0
  (3):\penalty0 266–285, 2018.

\bibitem[Mnih et~al.(2016)Mnih, Badia, Mirza, Graves, Lillicrap, Harley,
  Silver, and Kavukcuoglu]{Mnih:2016}
Volodymyr Mnih, Adrià~Puigdomènech Badia, Mehdi Mirza, Alex Graves,
  Timothy~P. Lillicrap, Tim Harley, David Silver, and Koray Kavukcuoglu.
\newblock Asynchronous methods for deep reinforcement learning.
\newblock \emph{ICML}, 2016.

\bibitem[Mohassel and Zhang(2017)]{Mohassel:2017}
Payman Mohassel and Yupeng Zhang.
\newblock Secureml: A system for scalable privacy-preserving machine learning.
\newblock \emph{IEEE Symposium on Security and Privacy}, 2017.

\bibitem[nVidia(2019{\natexlab{a}})]{Cuda2:2019}
nVidia.
\newblock Cuda c programming guide, 2019{\natexlab{a}}.
\newblock URL
  \url{https://docs.nvidia.com/cuda/cuda-c-programming-guide/index.html}.

\bibitem[nVidia(2019{\natexlab{b}})]{Cuda:2019}
nVidia.
\newblock Cuda c best practices, 2019{\natexlab{b}}.
\newblock URL
  \url{https://docs.nvidia.com/cuda/cuda-c-best-practices-guide/index.html}.

\bibitem[Papernot et~al.(2016)Papernot, McDaniel, Goodfellow, Jha, Celik, and
  Swami]{Papernot:2016}
Nicolas Papernot, Patrick McDaniel, Ian Goodfellow, Somesh Jha, Z.~Berkay
  Celik, and Ananthram Swami.
\newblock Practical black-box attacks against machine learning.
\newblock \emph{ACM Asia Conference on Computer and Communications Security},
  2016.

\bibitem[Pascanu et~al.(2015)Pascanu, Stokes, Sanossian, Marinescu, and
  Thomas]{Pascanu:2015}
Razvan Pascanu, Jack~W. Stokes, Hermineh Sanossian, Mady Marinescu, and Anil
  Thomas.
\newblock Malware classification with recurrent networks.
\newblock \emph{ICASSP}, 2015.

\bibitem[PRALab and s.r.l.(2019)]{SecML:2019}
PRALab and Pluribus~One s.r.l.
\newblock Secml library, 2019.
\newblock URL \url{https://gitlab.com/secml/secml}.

\bibitem[Radford et~al.(2019)Radford, Wu, Child, Luan, Amodei, and
  Sutskever]{Radford:2019}
Alec Radford, Jeffrey Wu, Rewon Child, David Luan, Dario Amodei, and Ilya
  Sutskever.
\newblock Language models are unsupervised multitask learners.
\newblock 2019.
\newblock URL \url{https://openai.com/blog/better-language-models/}.

\bibitem[Rapid7(2019)]{Metasploit:2019}
Rapid7.
\newblock Metasploit, 2019.
\newblock URL \url{https://www.metasploit.com}.

\bibitem[Ronen et~al.(2018)Ronen, Radu, Feuerstein, Yom-Tov, and
  Ahmadi]{Ronen:2018}
Royi Ronen, Marian Radu, Corina Feuerstein, Elad Yom-Tov, and Mansour Ahmadi.
\newblock Microsoft malware classification challenge.
\newblock \emph{arXiv:1802.10135}, 2018.

\bibitem[S.~Hettich(1999)]{KDD:1999}
S.~D.~Bay S.~Hettich.
\newblock Kdd cup 1999 data, 1999.
\newblock URL \url{http://kdd.ics.uci.edu/databases/kddcup99/kddcup99.html}.

\bibitem[Sanderson(2017)]{Sanderson:2017}
Grant~""3Blue1Brown"" Sanderson.
\newblock Deep learning, 2017.
\newblock URL \url{https://www.youtube.com/watch?v=aircAruvnKk}.

\bibitem[Sharif et~al.(2016)Sharif, Bhagavatula, Bauer, and
  Reiter]{Sharif:2016}
Mahmood Sharif, Sruti Bhagavatula, Lujo Bauer, and Michael~K. Reiter.
\newblock Accessorize to a crime: Real and stealthy attacks on state-of-the-art
  face recognition.
\newblock \emph{Conference on Computer and Communications Security}, 2016.

\bibitem[Shen et~al.(2016)Shen, Tople, and Saxena]{Shen:2016}
Shiqi Shen, Shruti Tople, and Prateek Saxena.
\newblock Auror: defending against poisoning attacks in collaborative deep
  learning systems.
\newblock \emph{32nd Annual Conference on Computer Security Applications},
  2016.

\bibitem[Shone et~al.(2018)Shone, Ngoc, Phai, and Shi]{Shone:2018}
Nathan Shone, Tran~Nguyen Ngoc, Vu~Dinh Phai, and Qi~Shi.
\newblock A deep learning approach to network intrusion detection.
\newblock \emph{IEEE Transactions on Emerging Topics in Computational
  Intelligence}, 2\penalty0 (1), 2018.

\bibitem[Spetlik and Razumenic(2019)]{Spetlik:2019}
Radim Spetlik and Ivan Razumenic.
\newblock Iris verification with convolutional neural network and unit-circle
  layer.
\newblock \emph{GCPR}, 2019.

\bibitem[Stevens et~al.(2017)Stevens, Suciu, Ruef, Hong, Hicks, and
  Dumitraş]{Stevens:2017}
Rock Stevens, Octavian Suciu, Andrew Ruef, Sanghyun Hong, Michael Hicks, and
  Tudor Dumitraş.
\newblock Summoning demons: The pursuit of exploitable bugs in machine
  learning.
\newblock \emph{arXiv:1701.04739}, 2017.

\bibitem[Sutskever et~al.(2014)Sutskever, Vinyals, and Le]{Sutskever:2014}
Ilya Sutskever, Oriol Vinyals, and Quoc~V. Le.
\newblock Sequence to sequence learning with neural networks.
\newblock \emph{NIPS}, 2014.

\bibitem[Sutton and Barto(2018)]{Sutton:2018}
Richard~S. Sutton and Andrew~G. Barto.
\newblock \emph{Reinforcement Learning: An Introduction}.
\newblock MIT Press, Cambridge, 2018.

\bibitem[Szegedy et~al.(2015)Szegedy, Liu, Jia, Sermanet, Reed, Anguelov,
  Erhan, Vanhoucke, and Rabinovich]{Szegedy:2015}
Christian Szegedy, Wei Liu, Yangqing Jia, Pierre Sermanet, Scott Reed, Dragomir
  Anguelov, Dumitru Erhan, Vincent Vanhoucke, and Andrew Rabinovich.
\newblock Going deeper with convolutions.
\newblock \emph{CVPR}, 2015.

\bibitem[Takaesu(2018)]{Takaesu:2018}
Isao Takaesu.
\newblock Deepexploit, 2018.
\newblock URL
  \url{https://github.com/13o-bbr-bbq/machine_learning_security/wiki}.

\bibitem[Takaesu(2019)]{Takaesu2:2019}
Isao Takaesu.
\newblock Security and machine learning, 2019.
\newblock URL
  \url{https://github.com/13o-bbr-bbq/machine_learning_security/tree/master/Security_and_MachineLearning}.

\bibitem[Tatoeba(2019)]{Tatoeba:2019}
Tatoeba.
\newblock Tatoeba project, 2019.
\newblock URL \url{https://tatoeba.org/}.

\bibitem[Xiao et~al.(2017)Xiao, Li, Zhang, and Xu]{Xiao:2017}
Qixue Xiao, Kang Li, Deyue Zhang, and Weilin Xu.
\newblock Security risks in deep learning implementations.
\newblock \emph{IEEE Symposium on Security and Privacy Workshops}, 2017.

\bibitem[Yamaguchi et~al.(1990)Yamaguchi, Sakamoto, Akabane, and
  Fujimoto]{Yamaguchi:1990}
Kouichi Yamaguchi, Kenji Sakamoto, Toshio Akabane, and Yoshiji Fujimoto.
\newblock A neural network for speaker-independent isolated word recognition.
\newblock \emph{ICSLP}, pages 1077--1080, 1990.

\bibitem[Yuan et~al.(2017)Yuan, He, Zhu, and Li]{Yuan:2017}
Xiaoyong Yuan, Pan He, Qile Zhu, and Xiaolin Li.
\newblock Adversarial examples: Attacks and defenses for deep learning.
\newblock \emph{arXiv:1712.07107}, 2017.

\bibitem[Zhou et~al.(2018)Zhou, Tang, Wang, Han, Liu, and Zhang]{Zhou:2018}
Zhe Zhou, Di~Tang, Xiaofeng Wang, Weili Han, Xiangyu Liu, and Kehuan Zhang.
\newblock Invisible mask: Practical attacks on face recognition with infrared.
\newblock \emph{arXiv:1803.04683}, 2018.

\bibitem[Zhu et~al.(2019)Zhu, Liu, and Han]{Zhu:2019}
Ligeng Zhu, Zhijian Liu, and Song Han.
\newblock Deep leakage from gradients.
\newblock \emph{NeurIPS}, 2019.

\end{thebibliography}

\end{document}